\documentclass[intlimits,twoside,a4paper]{article}

\usepackage{amsmath,amssymb}
\usepackage{epsfig}
\usepackage{graphicx}
\usepackage{wrapfig}
\usepackage{color}

\usepackage[T2A]{fontenc}
\usepackage[cp1251]{inputenc}

\usepackage{cmpj2}

\issue{2012}{15}{3}{33604}

\doinumber{10.5488/CMP.15.33604}

\title[OF-AIMD simulation study of the liquid noble metals]
{Orbital free ab initio molecular dynamics simulation study of some static and dynamic
properties\\ of liquid noble metals}

\author[G.M. Bhuiyan, L.E. Gonz\'alez, D.J. Gonz\'alez]{G.M. Bhuiyan\refaddr{label1}, L.E. Gonz\'alez\refaddr{label2}, D.J. Gonz\'alez\refaddr{label2}}

\addresses{
\addr{label1} Department of Theoretical
           Physics, University of Dhaka,
Dhaka--1000, Bangladesh
\addr{label2} Departamento de F\'\i sica Te\'orica, Universidad de Valladolid,
Valladolid, Spain
}

\date{Received June 19, 2012, in final form July 31, 2012}

\authorcopyright{G.M. Bhuiyan, L.E. Gonz\'alez, D.J. Gonz\'alez, 2012}

\begin{document}

\maketitle

\begin{abstract}
Several static and dynamic properties of liquid Cu, Ag and Au at thermodynamic
states near their respective melting points, have been evaluated by means of the
orbital free {\em ab-initio} molecular dynamics simulation method.
The calculated static structure shows good agreement with the available
X-ray and neutron diffraction data. As for the  dynamic properties, the calculated
dynamic structure factors point to the existence of collective density excitations along
with a positive dispersion for l--Cu and l--Ag.
Several transport coefficients have been obtained which show a reasonable agreement with
the available experimental data.
\keywords liquid noble metals, orbital free density functional theory, molecular
dynamics simulations, static structure, dynamic properties, transport coefficients
\pacs 61.25.Mv, 64.70.Fx, 71.15.Pd

\end{abstract}

\section{Introduction}
The {\it d}-electrons in the {\it d}-band metals are not so free as to justify a nearly free
electron (NFE) approach but, on the other hand, they are not so tightly  bound as
to be described by the tight binding method (TBM) or core electron theory.
Indeed, the study of $d$-band metals poses difficult theoretical challenges
although some progress has been made towards their understanding, both
in the solid and  liquid phases~\cite{Friedel,Pettifor,Sayers,Wills,Moriarty,Moriarty_2,Moriarty_3,Bretonnet,Hausleitner,Aryase,Dagens}.

From a theoretical point of view, accurate first principles electronic
structure calculations of {\it d}-band metals have been performed using the techniques
such as the linearized augmented plane wave or
the linearized muffin-tin orbital (LMTO) methods~\cite{Andersen}. Although
it is possible to  accurately calculate the interionic  forces within
these schemes~\cite{Soler,Yu}, still the computational demand of such calculations has
so far prevented its use within the context of Molecular Dynamics (MD) simulations.
As a consequence, most realistic
structural models for $d$-electron systems have been constructed by means
of empirical or semiempirical interatomic
potentials~\cite{Daw,Finnis,Foiles,Voter,Voter_2,Oh,BSS}.

In the particular case of the noble metals, the $d$-bands are completely filled but the
{\it sp-d} hybridization is still there~\cite{Wills,Gelatt}. This {\it sp-d}
hybridization effect can
be accounted for by either changing the {\it s, p, d} band occupancy number (in the case
of {\it ab-initio} pseudopotential theory) or by  using an effective
valence $Z$~\cite{Phuong}. In this respect, it has already been found
from the density functional based generalized pseudopotential theory that
the effective  {\it sp}-electron  valence lies~\cite{Moriarty,Moriarty_2,Moriarty_3} within the
range $1.1$ to $1.7$, where this non-integral number is mostly due to {\it sp-d}
hybridization effects~\cite{Moriarty,Moriarty_2,Moriarty_3,Gelatt,Phuong}.

The structure of the liquid noble metals has been studied at several temperatures by
Waseda~\cite{WasedaBook} using X-ray (XR)
diffraction methods. Neutron diffraction has been also used
in the case of Cu at two temperatures~\cite{Eder} and Ag near melting~\cite{Belli-Ag}.
As concerns their thermophysical properties, the situation is different in the case of
l--Cu and l--Ag on the one hand, and l--Au on the other hand.
Two recent compilations of thermophysical properties of liquid metals, due to Blairs~\cite{Blairs} and Singh {et al.}~\cite{Singh}, the latter including several temperatures,
analyze the previous experimental measurements
of the adiabatic sound velocity ($c_{\mathrm{s}}$), density ($\rho$), and specific heat at a constant
pressure ($C_{\mathrm{P}}$) of the systems,  and therefrom deduce several other magnitudes,
such as the isothermal compressibility ($\kappa_{\mathrm{T}}$)
or the ratio of specific heats at a constant pressure and at a constant volume ($\gamma$).
Now, in the cases of l--Cu and l--Ag, several experimental measurements were available,
so an assessment was performed and recommended values were given by the authors~\cite{Blairs,Singh}. On the contrary, only a single experiment is available to
determine the sound velocity of l--Au, within a wider study of the Au--Co alloy~\cite{Au-sound}. Therefore, one should consider that the uncertainty in the thermophysical
data of l--Au is larger that for l--Cu and l--Ag.
Other transport properties of the liquid noble metals, such as self-diffusion coefficient ($D$),
or shear viscosity ($\eta$) are readily available~\cite{ShimojiBook2,IGBook}.
In particular, the  self-diffusion coefficients of l--Cu over a wide temperature
range, have recently been determined by means of quasielastic
neutron scattering
measurements~\cite{Meyer}. More specifically,
the experimental data were used to calculate the self
intermediate scattering functions, $F_\mathrm{s}(q, t)$, at several 
$q$-values, and
the associated self-diffusion coefficients were evaluated from their decay rate
at small wavevectors.

Most theoretical
studies on the liquid noble metals have focused on
the static structural properties and thermodynamic properties, usually characterizing the liquid system by effective interatomic potentials constructed
either empirically by fitting to some experimental data or derived
from some approximate theoretical model. Therefrom, the liquid structure is
determined by resorting to either liquid state theories~\cite{HMcD} or to classical
molecular dynamics (CMD) simulations.

Holender {et al.}~\cite{Holender} have used the embedded atom model (EAM) to obtain
some effective interatomic potentials which were later on used
in CMD simulations aimed at evaluating the static structure of liquid noble metals
near melting.
Bogicevic {et al.}~\cite{Bogicevic} used the effective medium theory
to obtain a many-body potential
which, combined with CMD simulations, provided information on the static
properties and the self-diffusion coefficient of l--Au at different
temperatures. Their calculated pair distribution function, $g(r)$,
near melting has the main peak which is somewhat lower
than experiment and the subsequent oscillations are slightly out of phase.

Alemany {et al.}~\cite{Alemany1,Alemany2,Alemany3,Alemany4} used both
the EAM and TBM to derive many-body
potentials which were used in CMD simulations so as to obtain information
on various static and dynamic properties of l--Cu, l--Ag and l--Au.
Their calculated static structure factors, $S(q)$, showed a
good agreement with experiment except for a somewhat smaller height of the main peak.
They also obtained reasonable estimates for the
self-diffusion coefficients excepting l--Cu which was clearly underestimated.
A similar approach was used by Han {et al.}~\cite{Han} to evaluate the
self-diffusion and shear viscosity coefficients in liquid and undercooled Cu.
We also note that other workers~\cite{BSS,Bhuiyan2,Bhuiyan2_2} have
resorted to integral equation-type liquid state theories which, combined
with semiempirical interatomic potentials, have lead to reasonable estimates for
several static and thermodynamic
properties of a range of 3$d$, 4$d$ and 5$d$ liquid
transition metals.

In principle, an accurate approach to the study of the static and dynamic properties of the
liquid noble metals, would be provided by
{\it ab-initio} molecular dynamics (AIMD) simulation methods, which have
become widespread in the last twenty years or so.
Most AIMD methods are based on the density functional
theory (DFT)~\cite{HK,KS} which permits to calculate  the ground
state electronic energy of a collection of atoms, for given nuclear
positions, as well as yields the forces on the
nuclei via the Hellmann-Feynman theorem.  It enables one to perform
MD simulations in which the nuclear positions evolve according to
classical mechanics whereas the electronic subsystem follows adiabatically.
The Kohn-Sham (KS) orbital representation of the DFT (KS-AIMD method) has been
the usual approach when performing AIMD simulations although it is acknowledged that
this approach imposes heavy computational demands which limit
the size of the systems 
as well as the simulation times.
These limitations are enhanced in the case of $d$-electron systems such as the noble and
transition metals because a large number of electronic orbitals are needed.
Nevertheless, and despite the
above shortcomings, a few AIMD studies have already been performed on the liquid noble
metals~\cite{Pasquarello,Mitrohkin,Ganesh,chinos,Jakse-Au}.

The first AIMD calculation of l--Cu was performed by 
Pasquarello {et al.}~\cite{Pasquarello}, who studied some
static properties 
near melting using ultrasoft pseudopotentials~\cite{Vanderbilt} combined with a
plane-wave expansion for the electronic orbitals.
The simulation used 50 atoms, lasted for 2~ps and results were
obtained for the pair distribution function, the self-diffusion coefficient and the
electronic density of states.
More recently, Mitrohkin~\cite{Mitrohkin} has performed AIMD simulations
to analyze the melting process in Cu. The study used 62 atoms, lasted for 3~ps and produced
results for some static properties and diffusion coefficient of l--Cu near melting.
Two further AIMD studies of Cu~\cite{Ganesh,chinos}
focused on the possible appearance of
icosahedral arrangements of atoms in liquid and undercooled Cu, with sample
sizes between 100 and 200 particles and equilibrium
simulation times from 1 to 5~ps.
Pasturel {et al.}~\cite{Jakse-Au}, within a wider study of Au--Si alloys,
also performed AIMD simulations
of liquid and undercooled Au, using 256 atoms and equilibrium runs 6~ps long,
and obtained results for the temperature variation of the
self-diffusion coefficient and icosahedral atomic arrangements.
However, none of these AIMD calculations produced results for the dynamical
properties, because its evaluation requires in general larger systems, and,
in particular, substantially longer simulation times.

This goal can be achieved by resorting to the orbital free ab-initio molecular
dynamics (OF-AIMD) simulation method~\cite{Pearson,Pearson_2,Pearson_3,Pearson_4,GGLS,GGLS_2,BhuiyanSn}.
It is based on the Hohenberg and Kohn version of the
DFT theory~\cite{HK} where the electronic orbitals are replaced by
the total valence electron density which now becomes the basic variable.
This procedure greatly reduces the number of variables
describing electronic states and, therefore, it enables one to study larger
samples (a few thousand atoms) and for longer simulation times (tens of picoseconds).
Now, the interaction among the positive ions and the valence
electrons is characterized by means of a local pseudopotential which plays an
important role in  determining the  ground state energy and the
realistic forces acting  on ions.

This paper reports an OF-AIMD study of the static and dynamic properties of the
liquid noble metals (Cu, Ag, Au) at thermodynamic conditions close to their
respective melting points.
The layout of the paper is as follows. In section~2 we
briefly describe the OF-AIMD method and provide some technical
details. We also describe the local ionic pseudopotentials used in this calculations.
Section~3 reports and discusses the results
of the ab-initio simulations for several static and dynamic
properties which, moreover, are compared with the available
experimental data.
We conclude this paper in section~4.


\section{Theory}

A simple liquid metal is modelled as a disordered array of
$N$ bare ions with valence $Z$, enclosed in a volume $V$, and
interacting with $N_{\rm e}=NZ$ valence electrons through an electron-ion potential $v(r)$.
The total potential energy of the system can be written, within the
Born-Oppenheimer approximation, as the sum of the direct ion-ion coulombic
interaction energy and the ground state energy of the electronic system
under the external
potential created by the ions, $V_{\rm ext}
(\vec{r},\{\vec{R}_l\}) = \sum_{i=1}^N v(|\vec{r}-\vec{R}_i|)$ ,
\begin{equation}
E(\{\vec{R}_l\}) = \sum_{i<j} \frac{Z^2}{|\vec{R}_i-\vec{R}_j|} +
E_\mathrm{g}[n_\mathrm{g}(\vec{r}),V_{\rm ext}(\vec{r},\{\vec{R}_l\})] \, ,
\end{equation}
where $n_\mathrm{g}(\vec{r})$ is the ground state valence electron density and
$\vec{R}_l$ are the ionic positions.
According to DFT, the ground state valence electron density, $n_\mathrm{g}(\vec{r})$,
can be obtained by minimizing an energy functional $E[n]$, which can be written
\begin{equation}
E[n(\vec{r})] =
T_{\mathrm{s}}[n]+ E_H[n]+ E_{\rm xc}[n]+ E_{\rm ext}[n] \, ,
\label{etotal}
\end{equation}
where the terms represent, respectively,
the electronic kinetic energy, $T_{\mathrm{s}}[n]$,
of a non-interacting system of density $n(\vec{r})$,
the classical electrostatic energy (Hartree term),
the exchange-correlation
energy, $E_{\rm xc}[n]$, for which we used the local
density approximation and
finally the electron-ion interaction energy,
$E_{\rm ext}[n]$, where the electron-ion potential is
characterized by a local ionic pseudopotential,
\begin{equation}
E_{\rm ext}[n] = \int \rd\vec{r} \, n(\vec{r}) V_{\rm ext}(\vec{r}) \, .
\end{equation}
For $T_{\mathrm{s}}[n]$ we used an explicit, albeit approximate, functional of the valence electron
density. Several expressions were proposed and in the present calculations we used an average density model~\cite{GGLS,GGLS_2}, which provided a good description for
a range of liquid simple metals, namely
$T_{\mathrm{s}}[n]=T_\mathrm{W} [n]+T_{\alpha}[n]$, where
\begin{equation}
T_\mathrm{W}[n(\vec{r})] = \frac18 \int \rd\vec{r} \,
|\nabla n(\vec{r})|^2 \left/n(\vec{r})\right.
\end{equation}
is the well-known von Weizs\"acker term, and
\begin{eqnarray}
T_{\alpha} [n] &=& \frac{3}{10} \int \rd\vec{r} \, [n(\vec{r})]^{5/3-2\alpha}
[\tilde{k}(\vec{r})]^2 \, ,\nonumber \\
\tilde{k}(\vec{r}) &=& (2k_\mathrm{F}^0)^3 \int \rd\vec{s} \, k(\vec{s})
w_{\alpha}\left(2k_F^0|\vec{r}-\vec{s}|\right)  \, ,
\end{eqnarray}
where $k(\vec{r})=(3\pi^2)^{1/3} \;  [n(\vec{r})]^{\alpha}$, $k_\mathrm{F}^0$ is the Fermi
wavevector for a mean electron density $n_\mathrm{e} = N_\mathrm{e}/V$, and $w_{\alpha}(x)$ is a
weight function chosen so that both the linear response theory and
Thomas-Fermi limits are correctly recovered. Further details
are given in reference~\cite{GGLS,GGLS_2}.

\begin{figure}[ht]
\begin{center}
\includegraphics[width=0.5\textwidth]{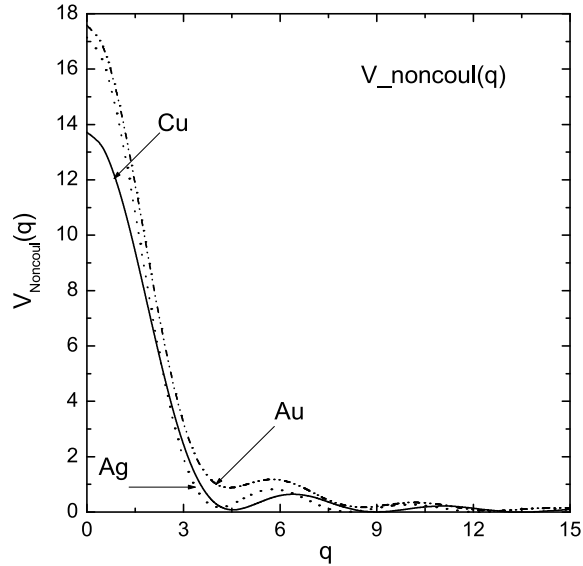}\hspace{2cm}
\end{center}
\caption{Non-Coulombic part of the electron-ion interaction for
liquid Cu, Ag and Au.}
\label{psfig}
\end{figure}
Another basic ingredient in the above formalism, is the local ionic
pseudopotential, $v_\mathrm{ps}(r)$, that describes the ion-electron interaction.
The AIMD  simulations based on KS-AIMD method usually employ
non-local pseudopotentials~\cite{Trouillers} obtained by fitting to some
properties of the free atom~\cite{Calderin,Calderin_2,Itami}. However, in the present
OF-AIMD  approach, the valence electron density is the basic variable,
and non-local pseudopotentials cannot be used. Therefore, the interaction among the
valence electrons and the ions must be described using a local pseudopotential which
 is usually chosen so as to include an accurate description of the electronic structure
in the physical state of interest. Bhuiyan {et al.}~\cite{BhuiyanSn} developed a local pseudopotential model which in conjunction with the
OF-AIMD method has provided a good description of several static and dynamic
properties of l--Sn near melting~\cite{BhuiyanSn}. Specifically, it is defined as
\begin{equation}
v_\mathrm{ps} (r)= \left\{
\begin{array}{rcl}
 A + B\,\exp(-r/a), & &  r<R_\mathrm{C}\, , \nonumber \\
 -Z/r, & & r > R_\mathrm{C}\, ,
\end{array}
\right.
\end{equation}
where $A$ and $B$ are contants, $R_\mathrm{C}$ is a core radius and $a$ is the softness
parameter. Aiming to reduce the number of free parameters,
we impose the condition that
the logarithmic derivatives of the potential inside and
outside the core are exactly the same at the core radius.
This permits to eliminate $B$ as a parameter,
and the successfulness of this approach can be judged by the capability of recovering the available experimental data.
The other parameters $A$, $a$ and $R_\mathrm{C}$, and the effective valence $Z$,
have been chosen so that the
OF-AIMD simulation reproduces the experimental static structure factor.
The values obtained herein are given in
table~\ref{TableInput}, where we notice that for a
given system the parameters remain constant for both thermodynamic states.
In figure~\ref{psfig}  we depicted the non-Coulombic  part of the
ionic pseudopotential
for l--Cu, l--Ag and l--Au. It shows that in the long wavelength limit ($q \to 0$), the value
of $v_\mathrm{ps}(q)$ is the largest for l--Au and the smallest for l--Cu.
Note also that the phase of oscillations is different for each system.

\begin{table}[!ht]
\caption{Input parameters used in the calculations; temperature $T$,
ionic number density $\rho$, amplitude in the core $A$,
softness parameter $a$, core radius $R_\mathrm{C}$ and the
effective ionic valence $Z$.}
\label{TableInput}
\vspace{2ex}
\begin{center}
\begin{tabular}{|c|l|l|l|l|l|l|}  \hline
System & $T$~(K)  &  $\rho$(\AA$^{-3}$)  & $A$ (au)  &  $a$ (au)  & $R_\mathrm{C}$ (au) &
$Z$ \\  \hline  \hline
Cu     & 1423  & 0.0755  & 0.05  &  0.3  &  1.40  & 1.35 \\
       & 1773  & 0.0728  & 0.05   & 0.3  &  1.40   & 1.35 \\  \hline
Ag     & 1273  &  0.0517 & 0.05   & 0.3  &  1.55  & 1.35 \\
       & 1673  &  0.0496  & 0.05   & 0.3  & 1.55   & 1.35 \\  \hline
Au     & 1423  &  0.0525  & 0.05  &  0.2  &  1.50  & 1.35 \\
       & 1773  &  0.0517  & 0.05   & 0.2  &  1.50   & 1.35 \\ \hline
\end{tabular}
\end{center}
\end{table}

We stress that the combination of the OF-AIMD method with local ionic pseudopotentials
has already provided accurate descriptions
of several static and dynamic properties for a range of bulk liquid
simple metals and binary alloys~\cite{GGLS,GGLS_2,BGGLS,BGGLS_2,BGGLS_3}.

\section{Results and discussion}

OF-AIMD simulations have been performed for l--Cu, l--Ag and l--Au
at two thermodynamic states near their respective triple points. Those states were chosen
due to the availability of experimental XR diffraction data~\cite{WasedaBook}.
Table~\ref{TableInput} gives additional information about thermodynamic states and other
input parameters used for the simulation.

The simulations were carried out using  500 particles in a cubic cell
with periodic boundary conditions and whose size was appropriate for
the corresponding experimental ionic number density.
Given the ionic positions at time $t$, the
electronic energy functional is minimized with respect to
$n({\vec{r}})$ represented by a single
{\it effective orbital}, $\psi(\vec{r})$, defined as
$n(\vec{r})=[\psi(\vec{r})]^2$. The orbital is expanded in plane waves
which are truncated at a cutoff energy, $E_{\rm Cut}=20.0$~Ryd.
The energy minimization with respect to the
Fourier coefficients of the expansion is performed
every ionic time step using a quenching method which results in the
ground state valence electron density and energy.
The forces on the ions are obtained from the electronic ground state
via the Hellman-Feynman theorem, and the ionic positions and velocities are
updated by solving Newton's equations, using the
Verlet leapfrog algorithm with a timestep of $6.0\cdot  10^{-3}$~ps.
Equilibration in the simulations lasted 10~ps. and the calculation of
properties was made by averaging over 150~ps.

In this study, we have evaluated several liquid static properties
(pair distribution function and static structure factor)
as well as various dynamic
properties, both single-particle ones (velocity autocorrelation function,
mean square displacement) and collective ones (intermediate scattering
functions, dynamic structure factors, longitudinal and transverse
currents).
The calculation of the time correlation functions (CF) was performed by
taking time origins every five time steps. Several CF
have also a dependence on the wave vectors $\vec{q}$ which depend  only on $ q \equiv | \vec{q} |$ because  our system is
isotropic.

\subsection{Static Properties}

\subsubsection{Liquid Cu}

\begin{figure}[!t]
\centerline{
\includegraphics[width=6.5cm,clip]{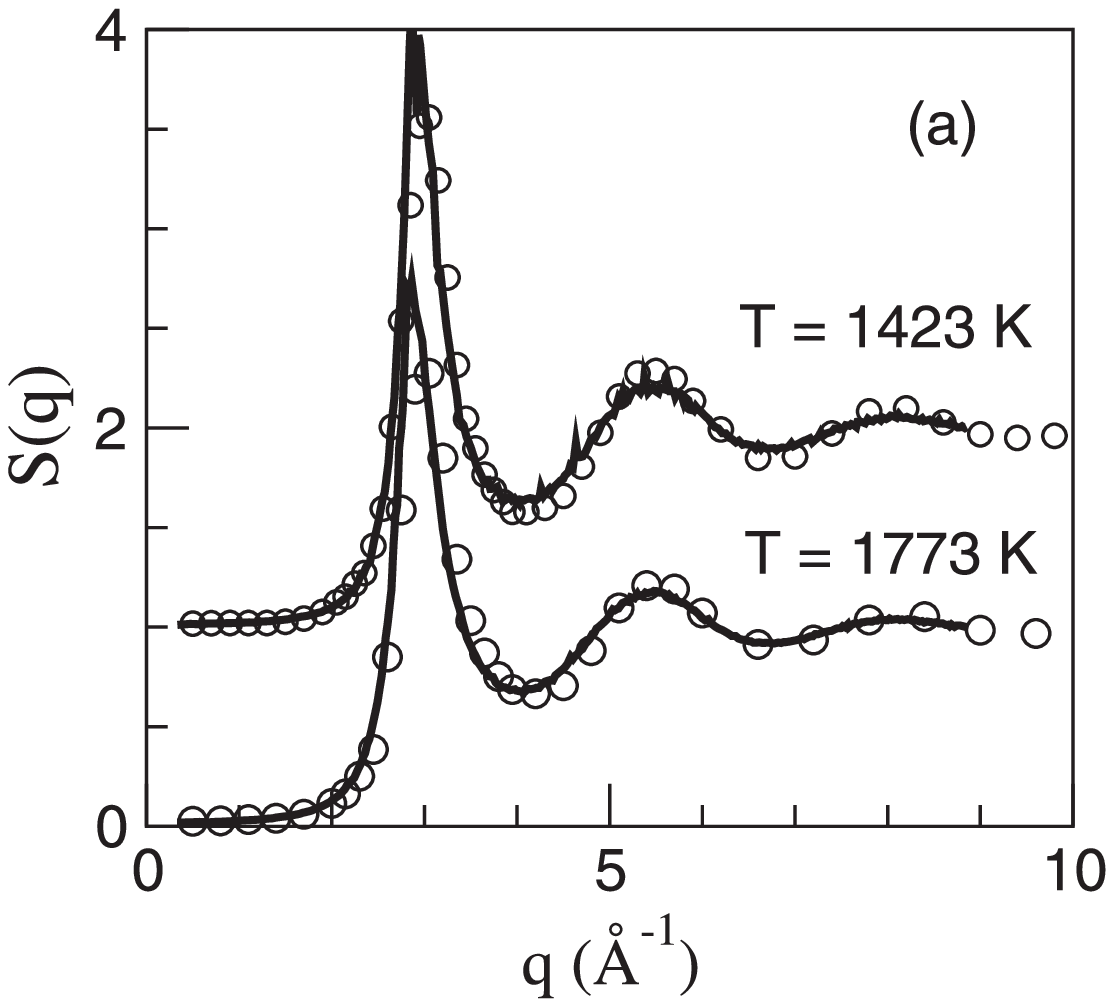}\hspace{8mm}\includegraphics[width=6.5cm,clip]{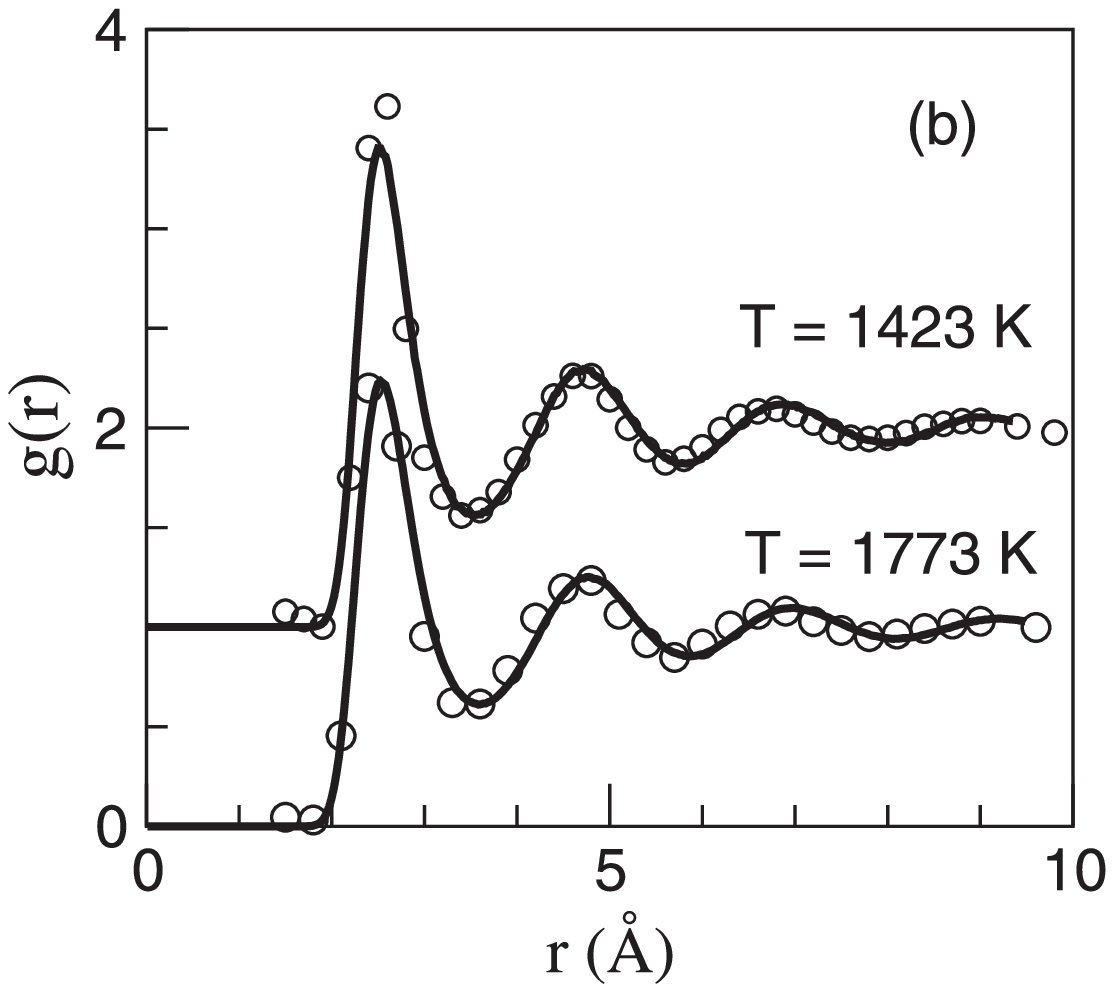}
}
\caption{(a) Static structure factors and (b) pair correlation functions for
l--Cu at two thermodynamic states. Solid lines are the OF-AIMD results and the open circles
stand for the XR diffraction data.}
\label{sqfig--Cu}
\end{figure}

The OF-AIMD simulation permits to directly evaluate  the static  structure
factor, $S(q)$, and its real space counterpart, i.e., the pair distribution function $g(r)$.
Figure~\ref{sqfig--Cu}~(a) shows the calculated  $S(q)$
for \linebreak l--Cu at two  different thermodynamic  states
characterized by  temperatures $T=1423$ and 1773~K. For both states, the main peak is located at $q_\mathrm{p} \approx 2.88$~\AA$^{-1}$. Comparison with the XR
data~\cite{WasedaBook} shows
an overall good agreement for both the positions and phases of the oscillations, although the
present OF-AIMD results slightly overestimate the height of the main peak.
Note, however, that the height of the main peak in the neutron data of Eder {et al.}
at 1393~K (not shown) is substantially higher than in the XR data, being in better
agreement with our results.
A similar overestimation of the height of the main peak of $S(q)$ in
l--Cu, as compared to XR measurements, was also obtained in CMD studies
carried out using  EAM-based  potentials~\cite{Foiles,Alemany3,Arai}.
The KS-AIMD of Ganesh and Widom~\cite{Ganesh} at 1398~K also yield a structure
factor with a height of the main peak similar to our data and to the neutron
measurements.
The agreement of our high temperature results with experiment is the same, while in this case,
the XR structure factor at 1773~K and the corresponding neutron data at 1833~K agree
better with each other than at lower temperatures.

The long wavelength limit of the static structure factor, $S(q \to 0$),
is linked with thermodynamics through the relationship
 $S(q \to 0)=\rho\,k_\mathrm{B}\,T\,\kappa_{\mathrm{T}}$  where  $k_\mathrm{B}$ is Boltzmann's constant and
$\kappa_{\mathrm{T}}$ is isothermal compressibility.
A least squares fit of
$S(q)=s_0+s_2 q^2+s_4q^4$ to the calculated $S(q)$ for small $q$-values
yields an estimate
 $\kappa_{T, \rm OF-AIMD}=0.90\pm 0.03$
(in units of 10$^{-11}$ N$^{-1}$ m$^2$ )  for $T = 1423$~K, underestimating the
experimental value of 1.49~\cite{ShimojiBook1,Blairs}, or $1.41$~\cite{Singh}.
For $T=1773$~K we find $\kappa_{\mathrm{T}}=1.09\pm 0.03$, while the experimental value is $1.74$
(in the same units)~\cite{Singh}.

The calculated pair distribution functions, $g(r)$, are depicted in
figure~\ref{sqfig--Cu}~(b) along
with the corresponding XR data~\cite{WasedaBook}.
The main peak is located at $r_\mathrm{p} = 2.53$~\AA \ and $2.55$~\AA \
for  $T=1443$  and $1773$~K, respectively, which
agrees with the corresponding experimental data. A similar good agreement is found for
the positions and  the phase of oscillations of the subsequent peaks.
The only noticeable discrepancy concerns the height of the main peak which is slightly
underestimated by the present calculations. Nevertheless, we note that
a similar disparity is also reported in KS-AIMD studies~\cite{Pasquarello,Ganesh,chinos}.
The average number of nearest neighbors, also known as coordination number (CN), is obtained by
integrating the radial distribution function (RDF), $4\pi r^2 \rho g(r)$, up
to a distance $r_\mathrm{m}$ which is usually identified as the position of the
first minimum in either the RDF or the $g(r)$~\cite{Cusak,McGreevy}.
Both choices often lead to rather similar results and in what follows we report
the results obtained by integrating up to the first minimum of the RDF which  was  found at
$r_\mathrm{m} \approx 3.42$ and $3.44$~\AA\  for $T=1443$ and 1773 K, leading to values
CN $\approx$ 12.9 and 12.6, respectively. For comparison, we note that the KS-AIMD studies
at 1500~K produce CN $\approx 12.5$~\cite{Pasquarello}, $12.3$~\cite{Ganesh},
and $12.9$~\cite{chinos} using a bit  different integration limits.

\subsubsection{Liquid Ag}

\begin{figure}[!t]
\centerline{
\includegraphics[width=6.5cm,clip]{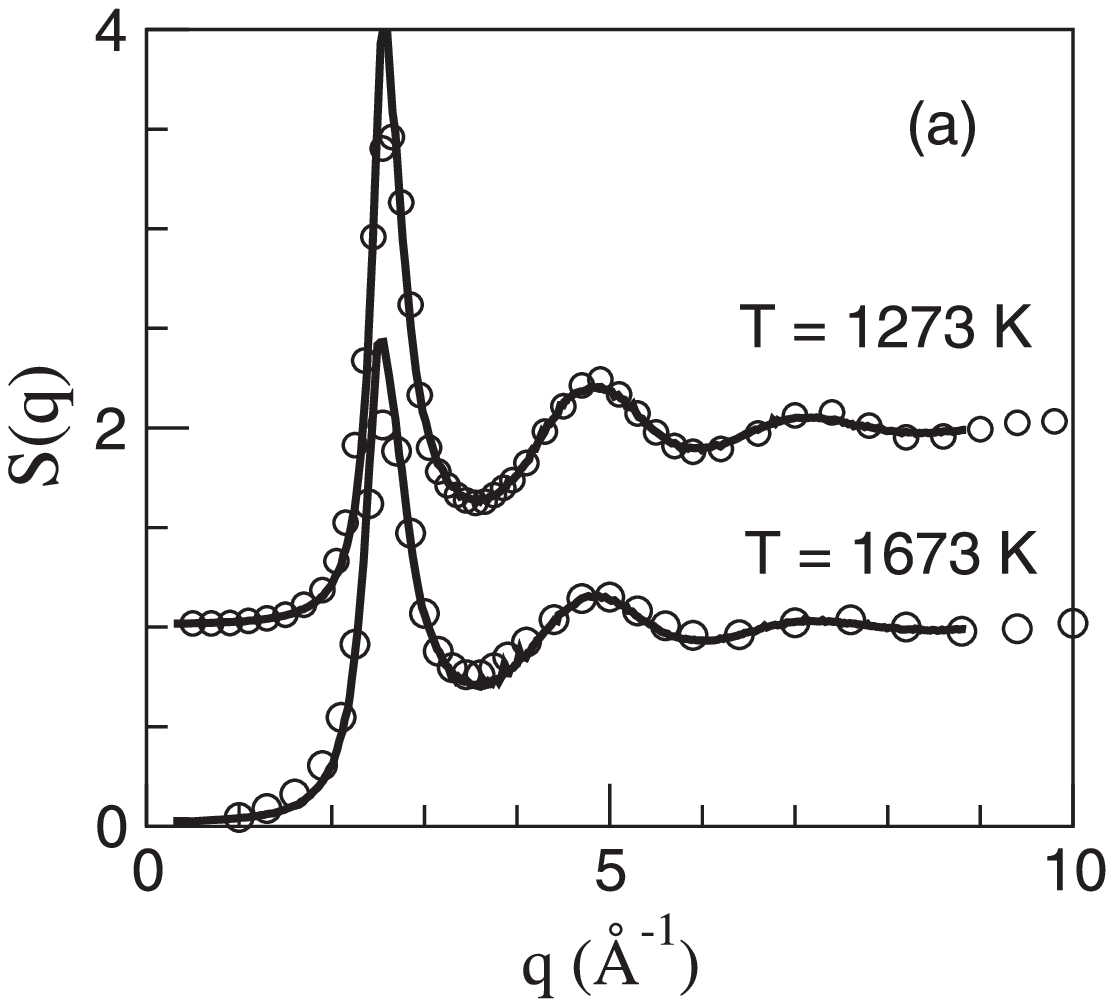}\hspace{8mm}
\includegraphics[width=6.5cm,clip]{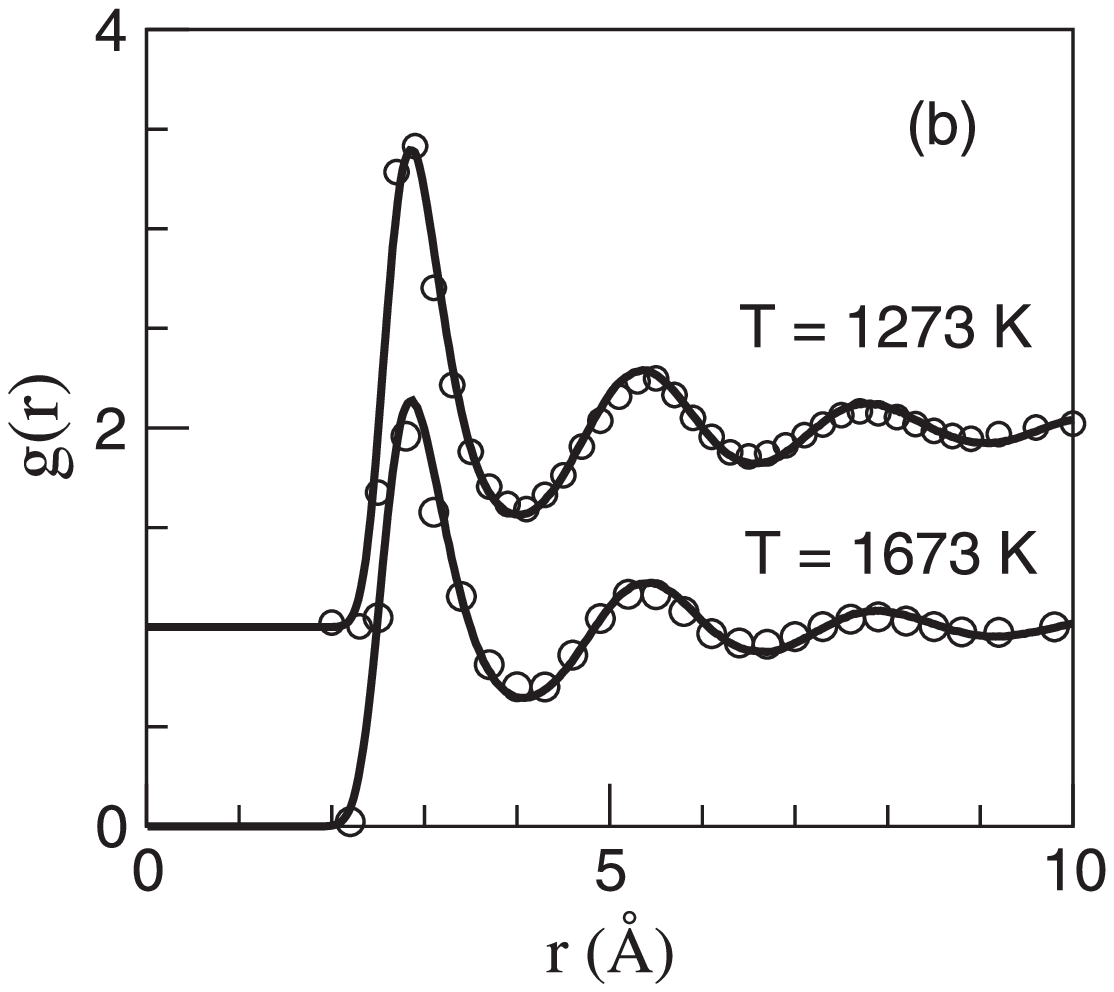}
}
\caption{(a) Static structure factors and (b) pair correlation functions for l--Ag at
two thermodynamic states.
Solid lines are the OF-AIMD results and the open circles are the XR diffraction data.}
\label{sqfig-Ag}
\end{figure}

The calculated $S(q)$ for l--Ag at two different
states are plotted in figure~\ref{sqfig-Ag}~(a) where they are compared with
the corresponding XR data~\cite{WasedaBook}.  The calculated position of the main peak
are at $q_\mathrm{p}=2.57$ and 2.59~\AA$^{-1}$ for $T=1273$~K and 1673~K, respectively.
For a lower temperature, $T=1273$~K, we observe that the calculated
height of the main peak is a bit bigger than that of the XR  data~\cite{WasedaBook};
indeed, a similar disparity has also been reported in other CMD studies
for l--Ag~\cite{Foiles,Alemany3}. Note also that the neutron $S(q)$ of Bellisent {et al.}
at 1323~K (not shown) has the height of the main peak of $2.85$, which is much higher than
Waseda's data, and is more in line with our result. On the other hand,
the positions and phase of oscillations of the subsequent
peaks are found to be in very good agreement with experiment.
We have also calculated the isothermal compressibility of
l--Ag at $T=1273$~K and we have obtained  $\kappa_{\mathrm{T}}=
1.94\pm 0.08$
(in units of $10^{-11}$ N$^{-1}$ m$^{2}$) to be compared with the  experimental
data of $2.11$~\cite{ShimojiBook1}, $1.92$~\cite{Blairs}, or $1.80$~\cite{Singh}.
For $T= 1673$~K, we have obtained $\kappa_{\mathrm{T}}=2.19\pm 0.05$, while experiment yields $2.21$~\cite{Singh}.

The  calculated pair correlation functions, $g(r)$, for l--Ag are depicted in
figure~\ref{sqfig-Ag}~(b)
for  $T= 1273$~K and 1673~K where we observe a good agreement
with the respective XR data~\cite{WasedaBook}.
Integrating up to the  first minima of the RDF,
found at $r_\mathrm{m}=3.86$~\AA \ and 3.82~\AA \ for
$T=1273$ and 1673~K, respectively, we obtain the values CN $\approx$
12.6 and 11.7, respectively.

\begin{figure}[!b]
\centerline{
\includegraphics[width=6.5cm,clip]{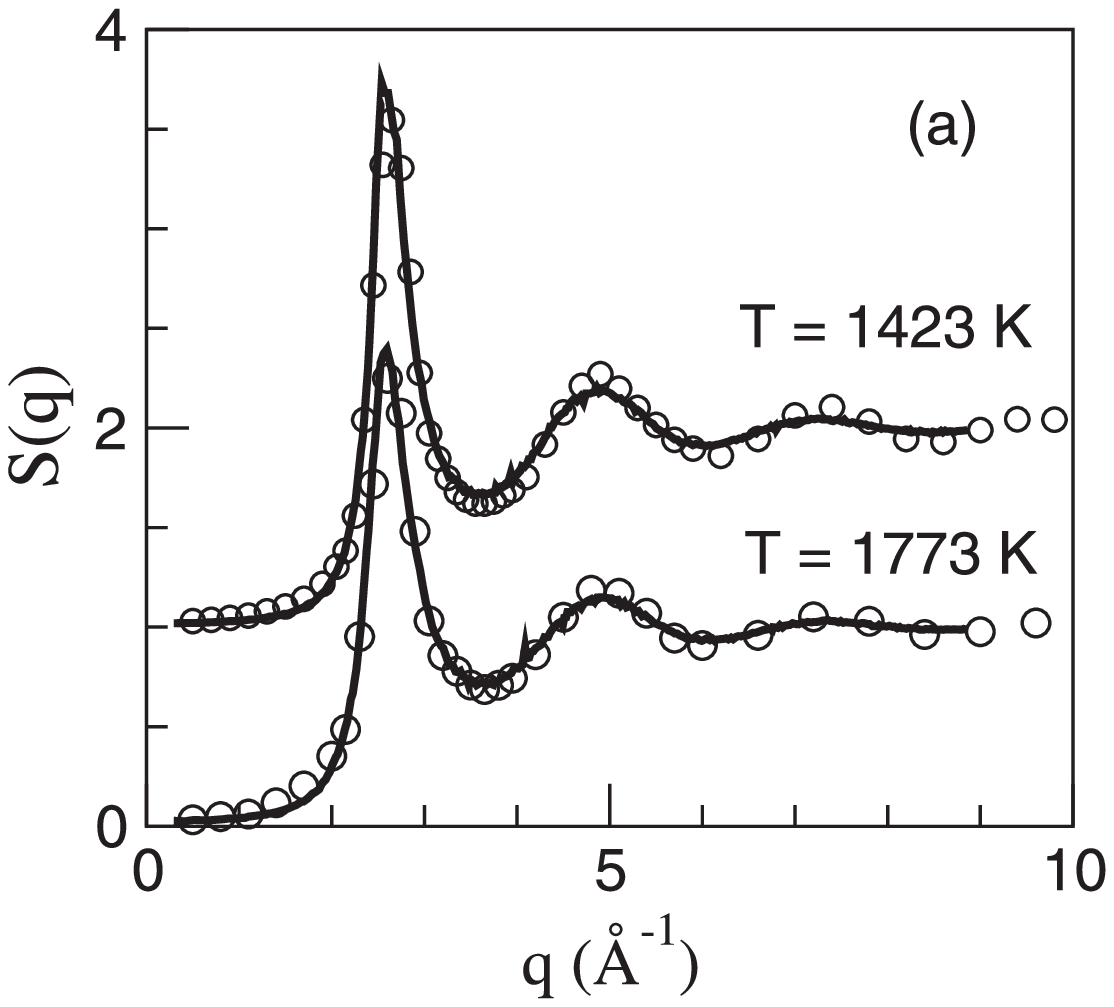}\hspace{8mm}
\includegraphics[width=6.5cm,clip]{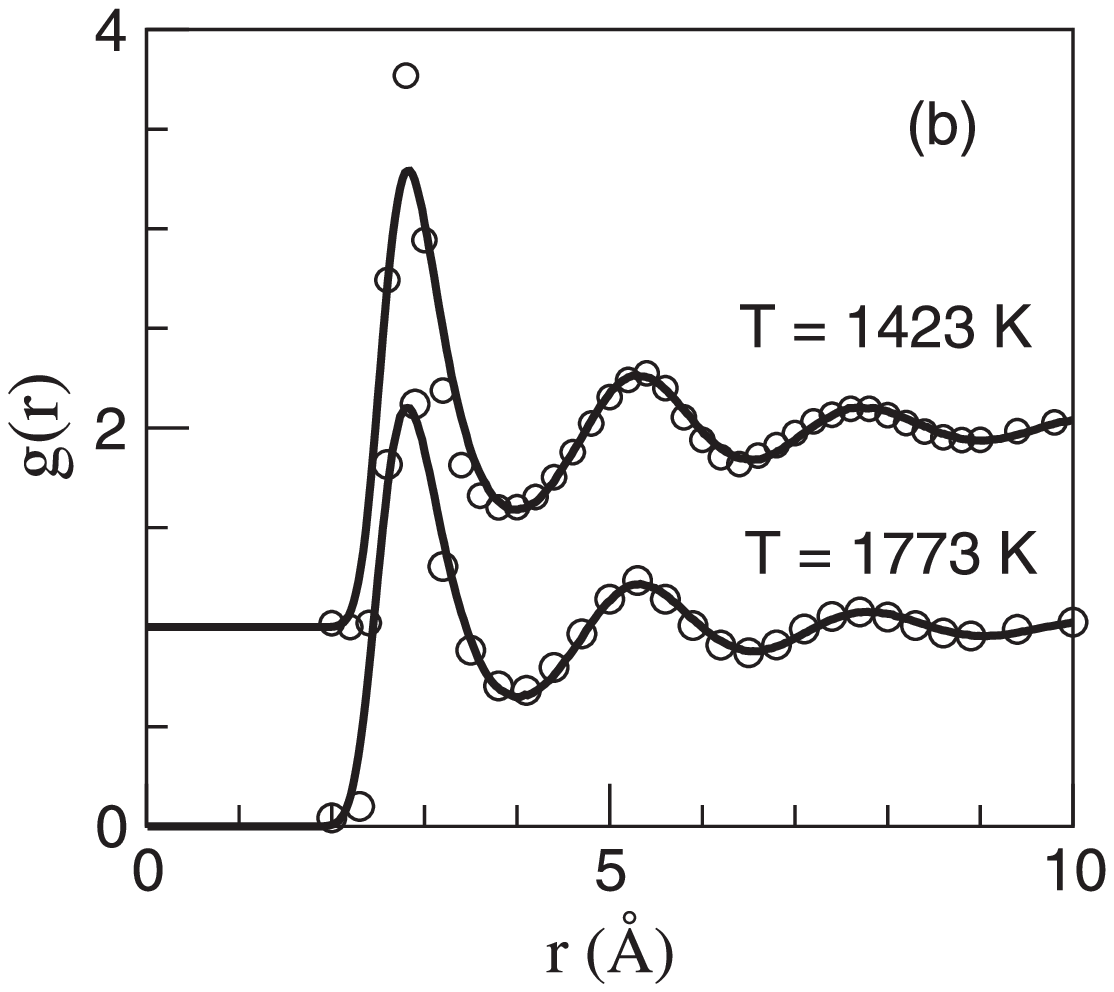}
}
\caption{(a) Static structure factors and (b) pair correlation functions for l--Au at
two thermodynamic states.
Solid lines are the OF-AIMD results and the open circles are the XR diffraction data.}
\label{sqfig-Au}
\end{figure}

\subsubsection{Liquid Au}

\looseness=-1The calculated $S(q)$ for l--Au  at $T= 1423$ and $1773$~K are depicted in figure~\ref{sqfig-Au}~(a) along with the corresponding XR
data~\cite{WasedaBook}. For both states, the main peak is located at
$q_\mathrm{p}=2.60$~\AA$^{-1}$ and a good agreement with experiment is observed for
the  positions and magnitudes of the main and subsequent peaks.
The calculated isothermal compressibility has yielded values
$\kappa_{\mathrm{T}}=1.61\pm 0.07$ (in units of 10$^{-11}$\, N$^{-1}$~m$^{2}$) at 1423~K, to compare with
$1.31$~\cite{Blairs} or $1.27$~\cite{Singh}, and $\kappa_{\mathrm{T}}=2.06\pm 0.06$ at 1773~K,
where Singh {et al.} report $1.61$~\cite{Singh}.

The $g(r)$ for l--Au at $T= 1423$~K and 1773~K are depicted in figure~\ref{sqfig-Au}~(b).
For both states, the main peak is located at $r_\mathrm{p}= 2.80$~\AA, which coincides
with the experimental value, although the height of the main peak is
somewhat underestimated, especially for the lower temperature.
The RDF has a first minimum at $r_\mathrm{m}=3.86$~\AA \  and  3.82~\AA \
which yields  values of CN $\approx$
12.7 and 12.2 for $T=1423$~K and 1773~K, respectively.

\subsection{Dynamic properties: Single particle dynamics}

Relevant information concerning the single particle dynamics can be derived from
several magnitudes and here we report our results obtained for some of those magnitudes.

The self-intermediate  scattering function, $F_\mathrm{s}(q,t)$, provides a detailed information
on  the single particle dynamic properties
over different length scales going
from hydrodynamic ($q\rightarrow0$) to  free particle
($q\rightarrow \infty$) limits. This is defined as
\begin{displaymath}
 F_\mathrm{s}(q,t)=\frac{1}{N} \left\langle \sum_{j=1}^{N} \exp\left[\ri\vec{q}\vec{R_{j}}(t+t_{0})\right]\,
\exp\left[-\ri\vec{q}\vec{R}_{j}(t_0)\right]\right\rangle \, ,
\end{displaymath}
where  $\langle \dots \rangle $ denotes the average over time origins and
wavevectors with the same module.
Closely connected to the $F_\mathrm{s}(q, t)$,  is the
velocity autocorrelation function (VACF)
of a tagged ion in the fluid, $Z(t)$, which can be
obtained as the $q \to 0$ limit of the first-order memory function of the
$F_\mathrm{s}(q, t)$ although in the present simulations it was calculated
from its definition
\begin{equation}
Z (t) = \langle \vec{v}_1(t) \vec{v}_1(0) \rangle
\left/ \langle v_1^2 \rangle\right. \, ,
\end{equation}
which stands for the normalized VACF. It provides information on the motion of
an atom inside the cage created by the
shell of nearest neighbors. Besides, its time integral leads to the
self-diffusion coefficient, $D$, namely
\begin{equation}
D= \frac{1}{\beta m} \int_0^{\infty} Z(t) \rd t\, ,
\label{DZ}
\end{equation}
where $\beta=1/(k_\mathrm{B}T)$. $D$ can also be obtained
from the slope of the mean
square displacement $\delta R^2(t)$
of a tagged ion in the fluid, as
\begin{equation}
D= \lim_{t \to \infty} \delta R^2(t)/6t=
\lim_{t\to\infty}\frac16 \frac{\rd\ \delta R^2(t)}{\rd t} \, , \qquad
\delta R^2(t) \equiv \langle | \vec{R}_1(t) - \vec{R}_1(0) |^2 \rangle \, .
\label{DR}
\end{equation}
In the present OF-AIMD calculations, both routes have led to practically the
same $D$ value.

\subsubsection{Liquid Cu}

\begin{figure}[!t]
\centerline{
\includegraphics[width=0.44\textwidth,clip]{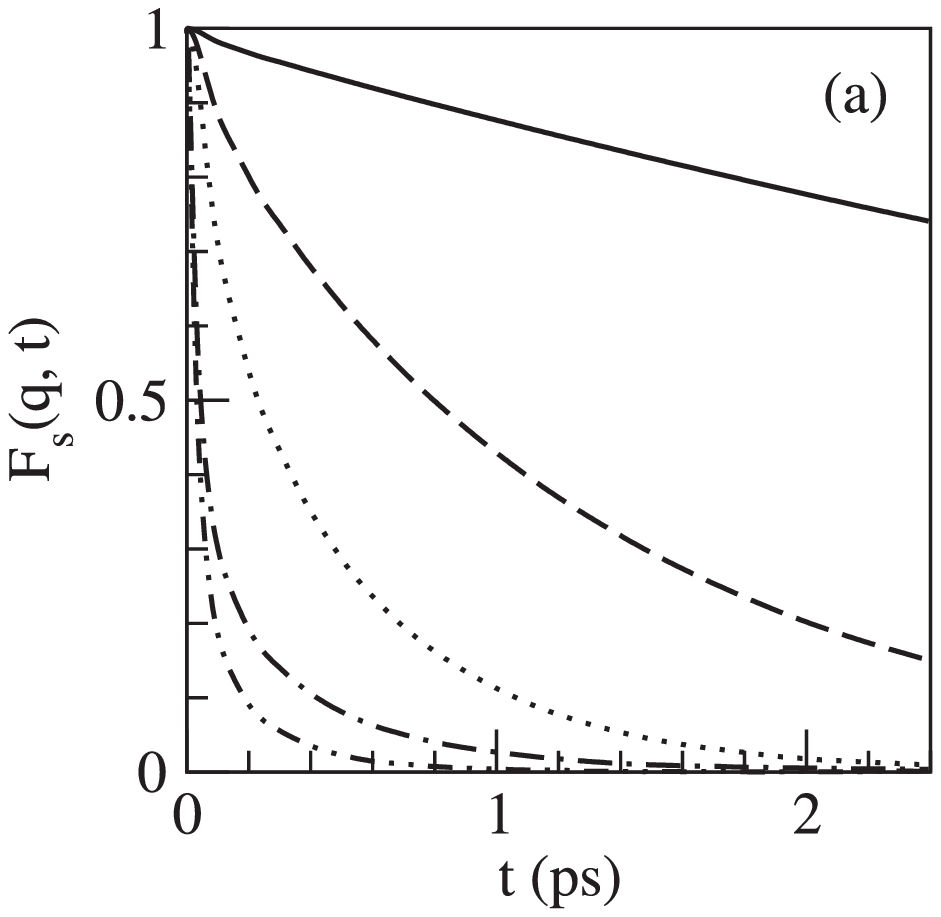}\hspace{8mm}
\includegraphics[width=0.46\textwidth,clip]{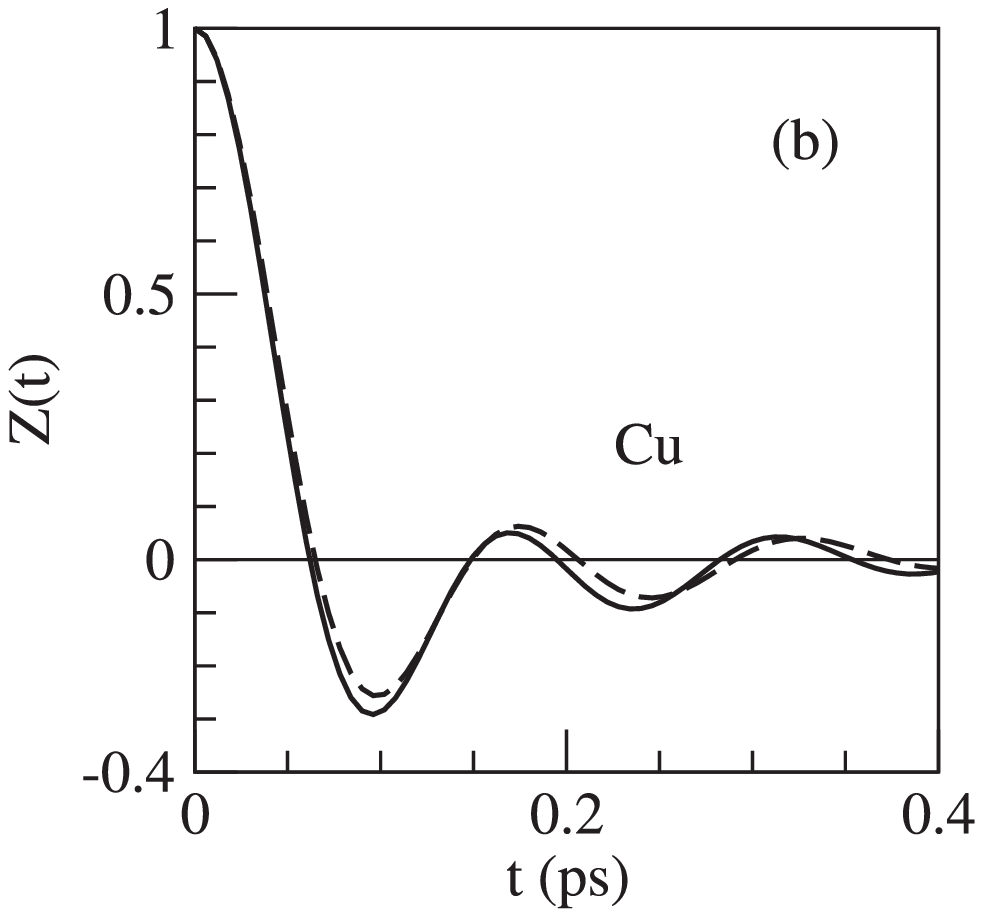}
}
\caption{(a) Self-intermediate scattering function of
l--Cu at $T=1423$~K.
Full line: 0.6~\AA$^{-1}$,
dashed line: 1.5~\AA$^{-1}$,
dotted line: 2.5~\AA$^{-1}$,
dotted-dashed line: 3.1~\AA$^{-1}$
and double dotted-dashed line: 4.3~\AA$^{-1}$
(b) Normalized velocity autocorrelation function
for l--Cu at 1423~K (full line) and 1773~K (dashed line).}
\label{Fsqt--Cu}
\end{figure}

Figure~\ref{Fsqt--Cu}~(a)
shows, for several $q$-values, the calculated
$F_\mathrm{s}(q,t)$ for l--Cu at $T=1423$~K. We observe the typical monotonous,
non-linear decrease with time which becomes faster with increasing
$q$-values; moreover, comparison with the simple liquid metals near their respective
melting points shows that at similar $q/q_\mathrm{p}$ values, the
$F_\mathrm{s}(q,t)$ has a comparable rate of decay~\cite{GGLS,GGLS_2,Balubook,Casas,Casas_2,Casas_3}.

The calculated $Z(t)$ for l--Cu are shown in figure~\ref{Fsqt--Cu}~(b).
The main features in the $Z(t)$ are comparable to those obtained for simple liquid
metals near melting~\cite{GGLS,GGLS_2,Balubook}, namely a first minimum about 0.30 deep
and a subsequent maximum with a rather weak amplitude. We recall that the negative
values of $Z(t)$ represent a backscattering effect induced by the cage effect;
moreover, with increasing temperature (and decreasing density) the cage effect
becomes less relevant, i.e., the first minimum in $Z(t)$ is shallower while the subsequent
oscillations are less marked.

\begin{figure}[!b]
\centerline{
\includegraphics[width=0.44\textwidth,clip]{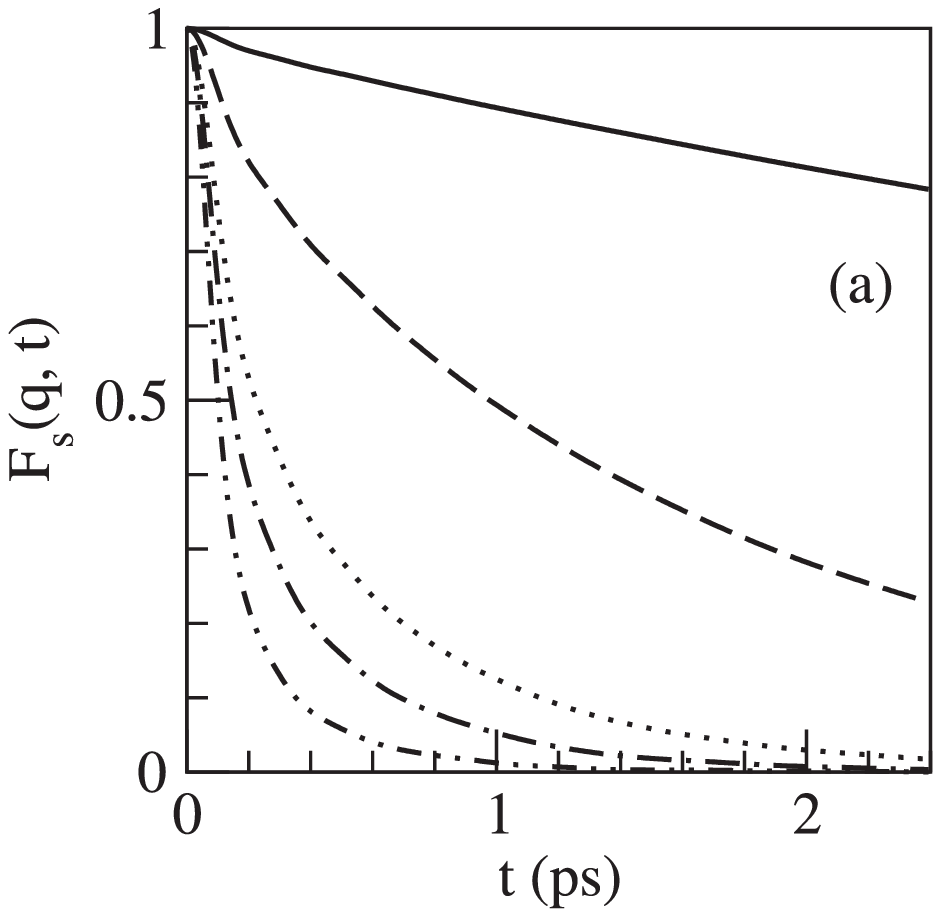}\hspace{8mm}
\includegraphics[width=0.46\textwidth,clip]{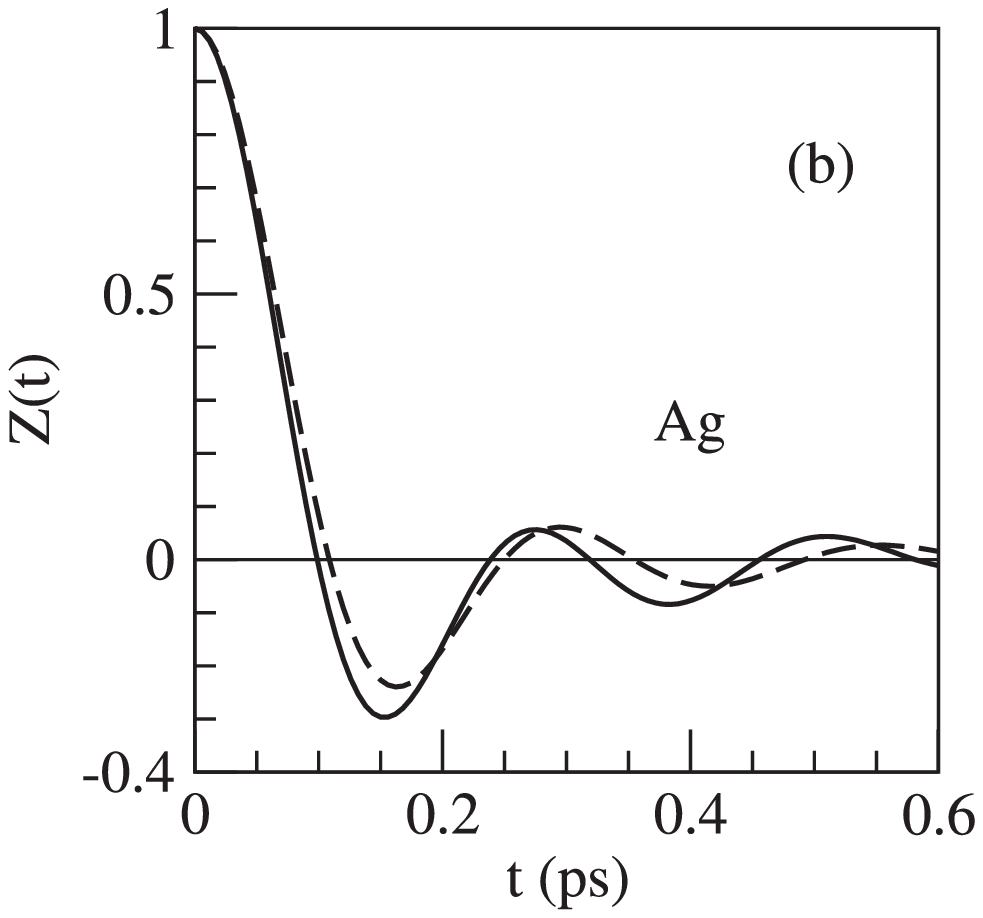}
}
\caption{(a) Self-intermediate scattering function of
l--Ag at $T=1273$~K.
Full line: 0.59~\AA$^{-1}$,
dashed line: 1.5~\AA$^{-1}$,
dotted line: 2.7~\AA$^{-1}$,
dotted-dashed line: 3.3~\AA$^{-1}$ and
double dotted-dashed line: 4.2~\AA$^{-1}$
(b) normalized velocity autocorrelation function for l--Ag at
1273~K (full line) and 1673~K (dashed line).}
\label{Fsqt-Ag}
\end{figure}

The self-diffusion coefficient was calculated according to equations~(\ref{DZ})--(\ref{DR}),  leading to values
$D = 0.39$~\AA$^{2}$/ps ( $T=1423$~K) and
0.58~\AA$^{2}$/ps ($T=1773$~K). The reference experimental
value at $T=1423$~K is 0.40~\AA$^{2}$/ps~\cite{ShimojiBook2,IGBook}, while
the recent measurements of Meyer yielded $D=0.37$~\AA$^2$/ps at 1420~K, which are both in
excellent agreement with our estimate.
Other CMD studies yielded
the values $D= 0.31$ and 0.27~\AA$^{2}$/ps ~\cite{Alemany3,Alemany4} and
$D= 0.36$~\AA$^{2}$/ps  for $T= 1400$~K~\cite{Han}.
The AIMD studies at 1500~K produced diffusion coefficients of
$D=0.28$~\cite{Pasquarello},
which clearly underestimates
the experimental data, presumably due to a small number of
particles (50 atoms) used in the simulation,
and $0.40\pm 0.05$~\cite{chinos} in good agreement with experiment.
The higher temperature is outside the range of measurements by Meyer~\cite{Meyer}
(up to 1620~K). However, he found that the measured data
could be well described through an Arrhenius formula. The value obtained with this
expression for $T=1773$~K is $D=0.65\pm 0.05$~\AA$^2$/ps, which is in reasonable agreement
with our result.
For this temperature, the CMD study of Han {et al.}~\cite{Han}
has reported $D = 0.59$~\AA$^{2}$/ps for $T=1700$~K, which is also similar
to our present estimate.

\subsubsection{Liquid Ag}

Figure~\ref{Fsqt-Ag}~(a) shows the calculated $F_\mathrm{s}(q,t)$,
at several $q$-values, for l--Ag at $T=1273$~K and we observe the
features  very similar to those already found in l--Cu.
The normalized VACF for l--Ag at $T=1273$  and $1673$~K are depicted in
figure~\ref{Fsqt-Ag}~(b) where we observe  the typical cage effect; now the variation with temperature is
more marked than in l--Cu because the relative change in the ionic density is greater.
Now the $Z(t)$ of l--Ag  becomes negative for longer times
and this is because the backscattering
associated with the cage effect in l--Ag is reduced by the combination of two factors, namely,
a smaller ionic number density and a greater atomic mass.
The calculated self-diffusion coefficients are $D=0.29$~\AA$^{2}$/ps
and $0.55$~\AA$^{2}$/ps for $T=1273$~K and $1673$~K, respectively, which
is very close to the experimental data of
$D=0.28$~\AA$^{2}$/ps and  0.58~\AA$^{2}$/ps~\cite{ShimojiBook2}

\subsubsection{Liquid Au}

\begin{figure}[!ht]
\centerline{
\includegraphics[width=0.44\textwidth,clip]{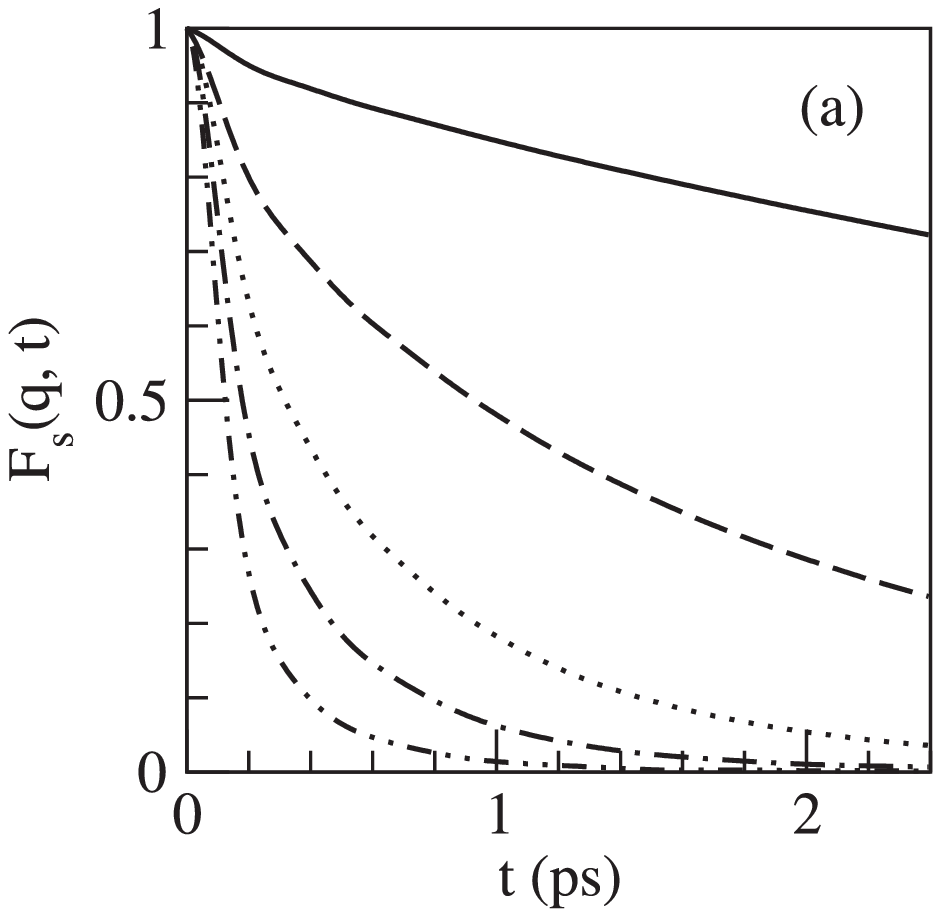}\hspace{8mm}
\includegraphics[width=0.46\textwidth,clip]{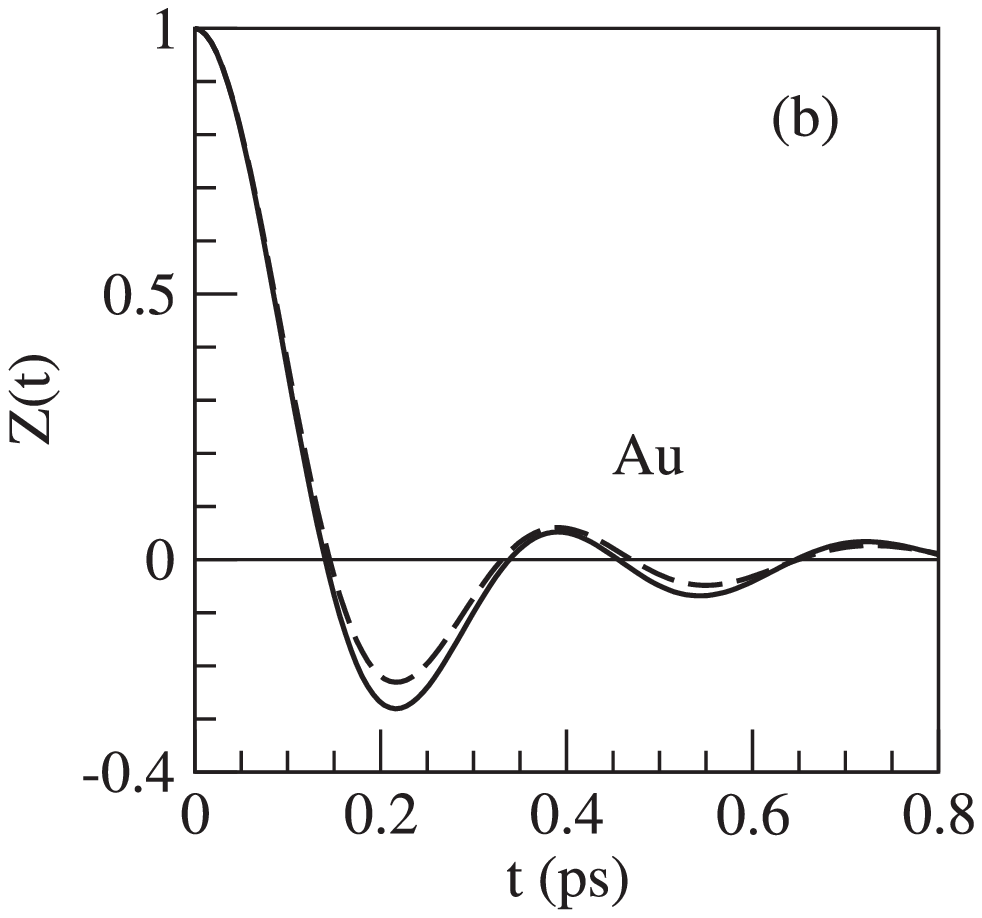}
}
\caption{(a) Self-intermediate scattering function of
l--Au at $T=1473$~K.
Full line: 0.59~\AA$^{-1}$,
dashed line: 1.39~\AA$^{-1}$,
dotted line: 2.41~\AA$^{-1}$,
dotted-dashed line: 3.2~\AA$^{-1}$ and
double dotted-dashed line: 4.2~\AA$^{-1}$
(b)~normalized velocity autocorrelation function for l--Au at
1423~K (full line) and 1773~K (dashed line).}
\label{Fsqt-Au}
\end{figure}

The calculated $F_\mathrm{s}(q,t)$,
at several $q$-values, for  l--Au at $T=1423$~K is depicted in
figure~\ref{Fsqt-Au}~(a).
As for the normalized VACF, figure~\ref{Fsqt-Au}~(b) shows the
calculated $Z(t)$ for $T=1423$  and $1773$~K.
Notice that in comparison with the previous results for l--Ag,
the $Z(t)$ for l--Au take longer to become negative and this is  due to a weaker
backscattering effect induced by the greater atomic mass
of the Au ions.
Comparison with the $Z(t)$  of Bogicevic {et al.}~\cite{Bogicevic} shows that
these authors obtain a $Z(t)$ with a narrower and shallower first minimum along with
weaker oscillations.
%
%
The calculated self-diffusion coefficients  are
$D = 0.27$ and 0.46~\AA$^{2}$/ps at $T=1423$ and 1773~K, respectively, to be compared with an
experimental value~\cite{ShimojiBook1} of 0.24~\AA$^{2}$/ps for l--Au at $T=1423$ K.
The calculations of Bogicevic {et al.}~\cite{Bogicevic} yielded somewhat
bigger values,
namely  $D= 0.31$ and 0.60~\AA$^{2}$/ps at $T=1423$ and $1773$~K, respectively,
whose origin can be traced back to the less marked cage effect in their $Z(t)$.
On the other hand, we mention that CMD simulations using two different EAM potentials
for l--Au at $T=1423$~K yielded $D= 0.26$~\AA$^{2}$/ps [43,48].
The AIMD simulations of
Pasturel {et al.}~\cite{Jakse-Au} produced a value of $D=0.153$~\AA$^2$/ps at 1400 K,
somewhat smaller than experiment, and $0.30$~\AA$^2$/ps at 1700 K.
To our knowledge, no experimental data are availble for the self-diffusion
coefficient of l--Au at $T=1773$ K.

\subsection{Dynamic properties: Collective dynamics}

Regarding the collective dynamics, the most important magnitude is the intermediate
scattering function, $F(q,t)$,
which provides information on the collective  dynamics of density fluctuations. It
is defined as
\begin{equation}
F(q,t)=\frac{1}{N}\,\left\langle \sum_{j=1}^{N}
\exp\left[\ri\bar{q}\vec{R_{j}}(t+t_{0})\right] \ \sum_{l=1}^{N}\exp\left[-\ri\vec{q}\vec{R_{l}}(t_{0})\right]\right\rangle .
\end{equation}
Its space Fourier transform (FT) produces the van Hove correlation function whereas its
time FT results in the dynamic structure factor, $S(q, \omega)$, which is
the magnitude
measured in the inelastic XR (or neutron) scattering experiments.

Another interesting magnitude associated with the density fluctuations is
 current due to the overall motion of particles, i.e.,
\begin{equation}
\vec{j}(q, t) =
\sum_{j=1}^{N} \vec{v}_{j}(t) \; \exp [ \ri \vec{q} \cdot
\vec{R}_{j}(t) ] \, ,
\label{jiqt}
\end{equation}
which is usually split into a longitudinal component,
$\vec{j}_\mathrm{L}(q, t)$ parallel to $\vec{q}$, and a transverse component,
$\vec{j}_\mathrm{T}(q, t)$, perpendicular to $\vec{q}$. Therefrom, the
longitudinal, $J_\mathrm{L}(q, t)$, and transverse $J_\mathrm{T}(q, t)$,
current correlation functions are obtained as
\begin{equation}
J_\mathrm{L}(q, t)= \frac{1}{N} \langle
\vec{j}_\mathrm{L}(q, t) \cdot \vec{j}^{ *}_\mathrm{L}(q, 0) \rangle  , \quad\quad
J_\mathrm{T}(q, t)= \frac{1}{2N} \; \langle
\vec{j}_\mathrm{T}(q, t) \cdot \vec{j}^{ *}_\mathrm{T}(q, 0) \rangle \, .
\label{CLTjqt}
\end{equation}
The corresponding time FT produce the associated spectra,
$J_\mathrm{L}(q, \omega)$ and $J_\mathrm{T}(q, \omega)$, respectively,
with \linebreak $J_\mathrm{L}(q, \omega) = \omega^2 S (q, \omega)$.
The transverse current  correlation
function $J_\mathrm{T}(q,t)$,  is not associated with any measurable quantity and can
only be determined by means of MD  simulations.
It provides information on the shear modes, and its shape evolves
from a Gaussian, in
both $q$ and $t$, for the  free  particle limit  towards a  Gaussian in $q$  and
exponential in $t$ for the hydrodynamic limit ($q\rightarrow0$), namely
\begin{equation}
 J_\mathrm{T}(q\rightarrow 0,t)=\frac{1}{\beta\,m}\,\exp\left[-q^{2}\eta|t|\left/(m\rho)\right.\right] \, ,
\end{equation}
where $\eta$ is the shear  viscosity.
On the other hand, for intermediate $q$-values, the $J_\mathrm{T}(q,t)$ shows a
complicated behaviour, because it may oscillate signaling the propagation of
shear waves. From the
calculated $J_\mathrm{T}(q,t)$ it is possible to obtain the
shear viscosity coefficient $\eta$ as follows. The memory function
representation of $J_\mathrm{T}(q,t)$,
\begin{equation}
\widetilde{J}_\mathrm{T}(q,z)=
\frac{1}{\beta\,m}\,\left[z+\frac{q^{2}}{\rho\,m}\widetilde{\eta}(q,z)\right]^{-1} \, ,
\end{equation}
where the tilde denotes the Laplace transform, introduces a generalized  shear
viscosity coefficient $\widetilde{\eta}(q,z)$. The area under the normalized
$J_\mathrm{T}(q,t)$ gives $\beta\,m \widetilde{J}_\mathrm{T}(q,z=0)$, from
which $\widetilde{\eta}(q,z=0) \equiv \widetilde{\eta}(q)$
can be obtained and when extrapolated to
$q \to 0$ it produces a usual shear viscosity coefficient $\eta$.
This is performed~\cite{Palmer} by exploiting the property that
inversion is a symmetry in the system and,
therefore,  $\widetilde{\eta}(q)$ should be an even
function of $q$ which permits to
approximate (when $q \to 0$) $\widetilde{\eta}(q)=\eta(1-\alpha\,q^{2})$.


\subsubsection{Liquid Cu}

\begin{figure}[!h]
\includegraphics[width=0.48\textwidth]{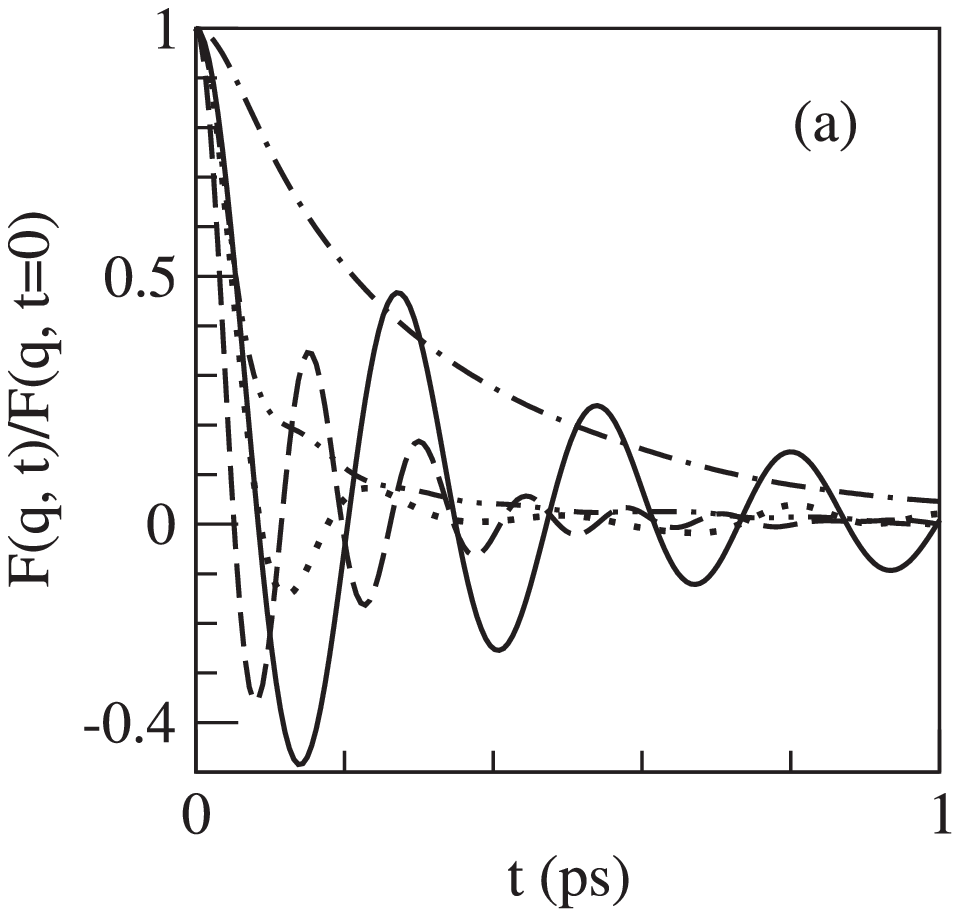}%
\hfill%
\includegraphics[width=0.48\textwidth]{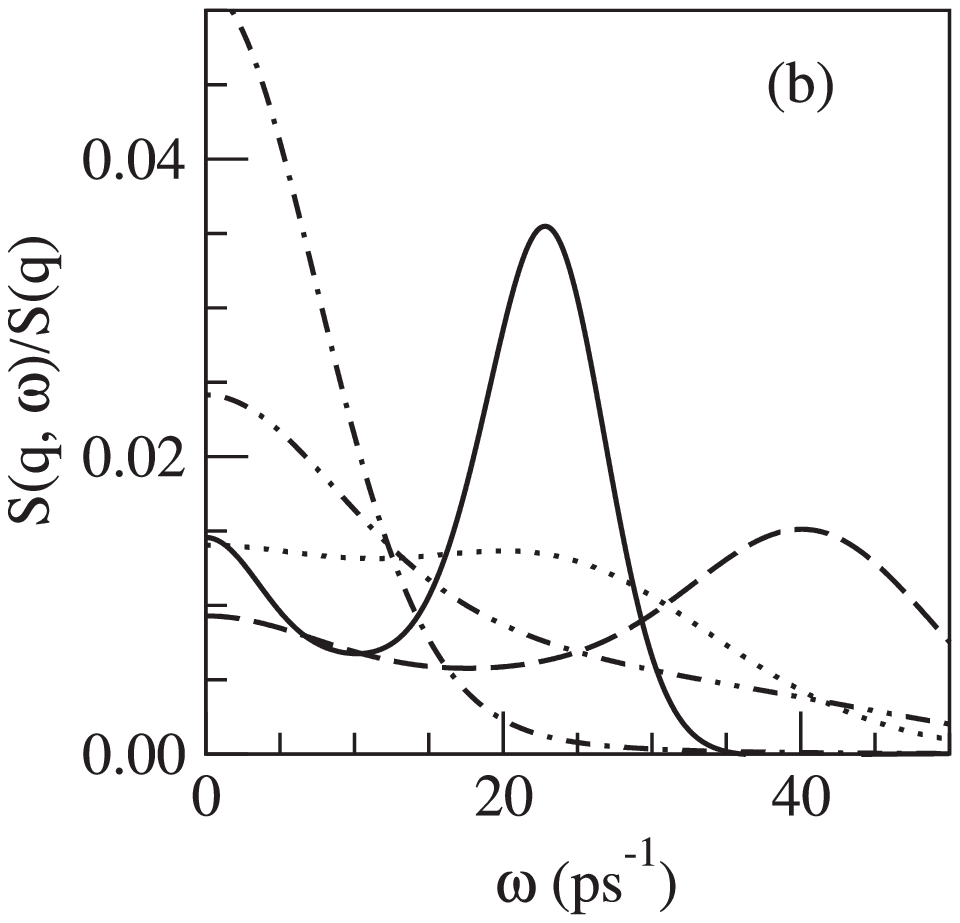}%
\\%
\parbox[t]{0.48\textwidth}{%
\centerline{(a)}%
}%
\hfill%
\parbox[t]{0.48\textwidth}{%
\centerline{(b)}%
}%
\caption{(a) Intermediate scattering function of
l--Cu at $T=1423$~K.
Full line: 0.58~\AA$^{-1}$,
dashed line: 1.5~\AA$^{-1}$,
dotted line: 2.22~\AA$^{-1}$,
dotted-dashed line: 3.1~\AA$^{-1}$,
and double dotted-dashed line: 4.31~\AA$^{-1}$
(b) Same as above but for the dynamic structure factor.}
\label{Fqt--Cu}
\end{figure}

Figure~\ref{Fqt--Cu}~(a) shows the calculated $F(q,t)/F(q,t=0)$ at
several $q$-values for l--Cu at $T=1423$~K. The $F(q,t)$
exhibit an oscillatory behaviour at low $q$-values with the oscillations becoming
weaker for increasing $q$'s until they finally disappear at
$q \approx$ 2.2~\AA$^{-1}$  $\approx 0.75 q_\mathrm{p}$. We stress that this behaviour is very
similar to the one found for the simple
liquid metals near melting~\cite{GGLS,GGLS_2,Balubook}.

\begin{figure}[!h]
\includegraphics[width=0.45\textwidth]{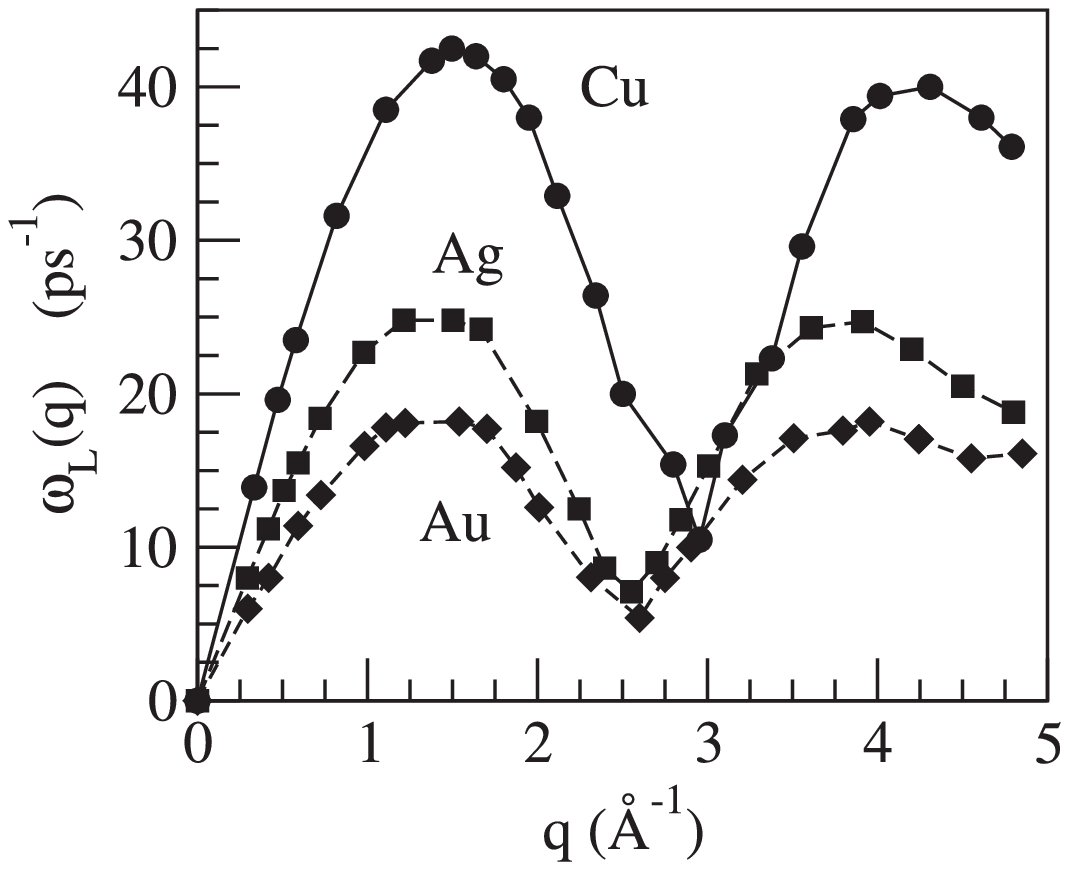}%
\hspace{27mm}%
\includegraphics[width=0.30\textwidth]{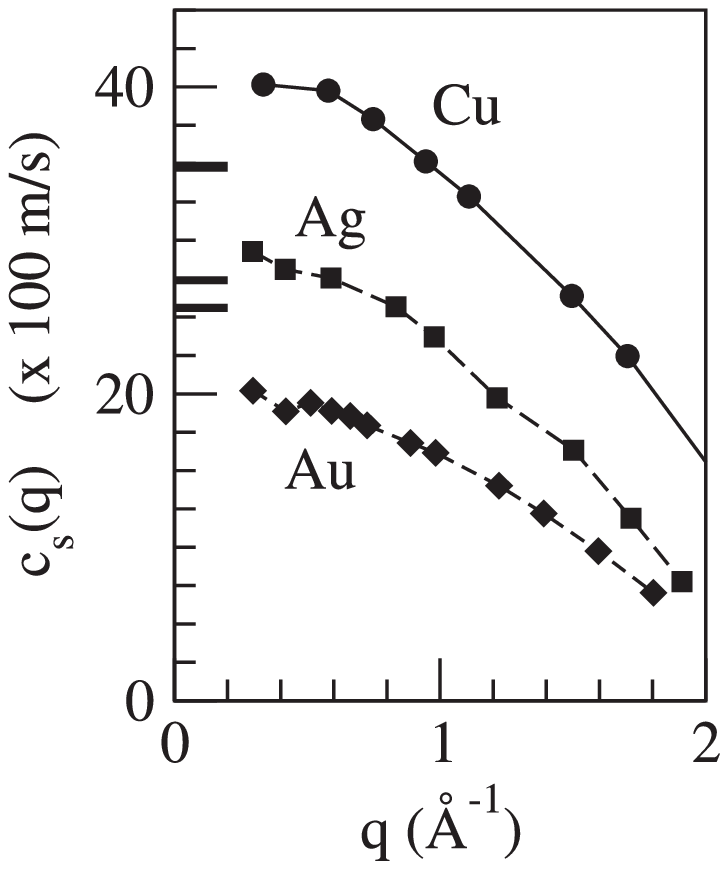}
\hspace{10mm}%
\\%
\parbox[t]{0.48\textwidth}{%
\caption{Dispersion relations from the peak positions of the calculated
$C_\mathrm{L}(q,\omega)=\omega^2 S(q, \omega)$
for l--Cu, l--Ag and l--Au at $T = 1423$, 1273  and 1423~K, respectively.}
\label{disperL}%
}%
\hfill%
\parbox[t]{0.48\textwidth}{%
\caption{$q$-dependent adiabatic sound velocity
for l--Cu, l--Ag and l--Au at $T = 1423$, 1273  and 1423~K, respectively. The full lines on the
$y$-axis stand for the respective hydrodynamic sound velocities.}
\label{Velq}%
}%
\end{figure}

From the calculated $F(q, t)$, we performed a time FT (with an
appropriate window to smooth out the truncation effects) leading to the associated
dynamic structure factor, $S(q, \omega)$. The obtained results are depicted in
figure~\ref{Fqt--Cu}~(b) for several $q$-values. We notice that up to
$q \approx 0.75 q_\mathrm{p}$, the calculated $S(q, \omega)$ exhibit
well defined side-peaks which are indicative of
collective density excitations.
The FT of the longitudinal current correlation function, $J_\mathrm{L}(q,\omega)$, shows,
however, side-peaks for all wavevectors, and from the
positions of these side-peaks, $\omega_\mathrm{L}(q)$, a dispersion relation of the
density fluctuations was obtained and plotted
in figure~\ref{disperL}.
In the hydrodynamic region (small $q$) the slope of
the dispersion  relation  curve is the  $q$-dependent adiabatic sound
velocity $c_{\mathrm{s}}(q)=v_\mathrm{th}\sqrt{\gamma/S(q)}$, with $v_\mathrm{th}=\sqrt{k_\mathrm{B}T/m}$ being
the thermal velocity, $\gamma$  the ratio of specific heats and $k_\mathrm{B}$ Boltzmann's
constant. In the $q\rightarrow 0$ limit, the $c_{\mathrm{s}}(q)$ reduces to the bulk
adiabatic sound velocity $c_{\mathrm{s}}$.
Using the smallest $q$-value achieved by the simulations,
$q_\mathrm{min}= 0.334 $~\AA$^{-1}$,
we get an estimate $c_{\mathrm{s}} (q_\mathrm{min}) \approx$ 3880~m/s, which is clearly above the
experimental hydrodynamic value $c_{\mathrm{s}} \approx$ 3481~\cite{Blairs} or 3449~m/s~\cite{Singh}.
In figure~\ref{Velq} we have plotted the
$c_{\mathrm{s}}(q)$  which clearly points to the existence of some positive dispersion  in l--Cu.
This behaviour qualitatively agrees with an estimate of 4230 m/s~\cite{Scopigno} obtained
with inelastic XR scattering data.
For the higher temperature $T=1773$~K, we get a smaller value  $c_{\mathrm{s}} \approx$ 3802~m/s,
whereas the experimental hydrodynamic sound velocity is 3266~m/s~\cite{Singh}, signaling
the persistence of the positive dispersion at these higher temperatures.

\begin{figure}[!t]
\centerline{
\includegraphics[width=16.pc,clip]{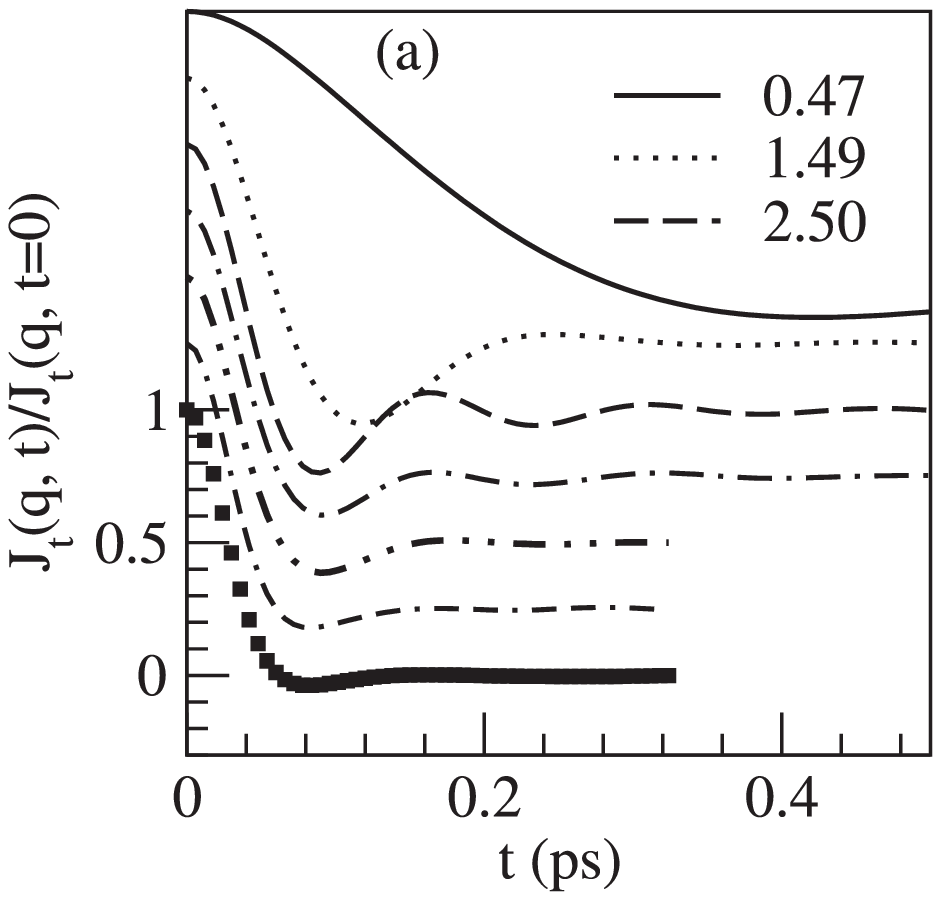}\hspace{8mm}
\includegraphics[width=16.pc,clip]{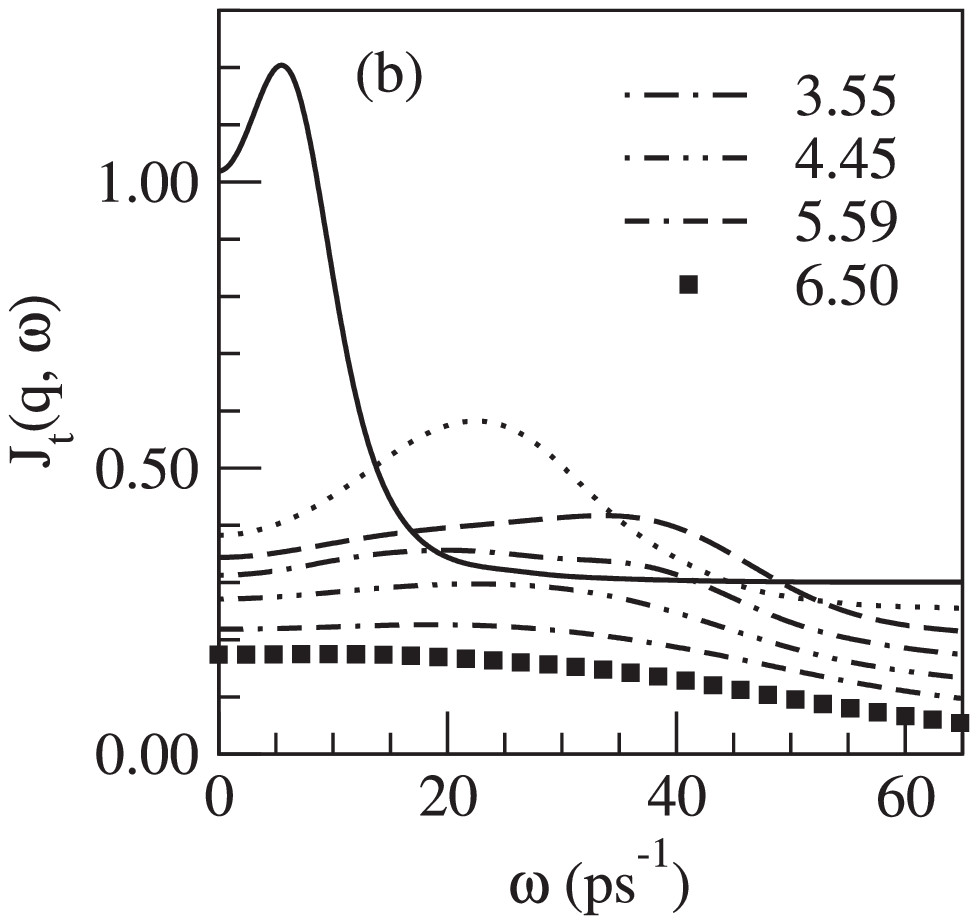}
}
\caption{(a) Transverse current correlation function of
l--Cu at $T=1423$~K. (b) Same as above but for the $J_\mathrm{T}(q, \omega)$.}
\label{Trans--Cu}
\end{figure}

Figure~\ref{Trans--Cu}~(a) shows the results for the normalized
$J_\mathrm{T}(q,t)$ of l--Cu at $T=1423$~K at several $q$ values.
Notice that for small $q$'s, the corresponding $J_\mathrm{T}(q,t)$ decrease slowly
but it becomes faster with increasing $q$ values.
The corresponding spectrum $J_\mathrm{T}(q,\omega)$ is depicted in
figure~\ref{Trans--Cu}~(b)
and  for some intermediate $q$-range we observe an inelastic peak
at nonzero frequency.
This peak, which reflects the propagation of shear waves in the liquid,
does not appear at the smallest value reached by the simulation
($q = 0.334$~\AA$^{-1}$) but it already shows up for
$q = 0.473$~\AA$^{-1}\approx0.16 q_\mathrm{p}$, and
remains up to  $q\approx 2.5 q_\mathrm{p}$.
The associated peak frequency increases with $q$, takes a maximum value at
$q \approx q_\mathrm{p}$, and then decreases with increasing $q$ as $J_\mathrm{T}(q, \omega)$
evolves towards a gaussian shape. In fact, we recall that a
similar behaviour has already been reported for the alkali metals~\cite{Balubook}
where the inelastic peak appears for $q \geqslant 0.07 q_\mathrm{p}$.
On the other hand, from the position of the peaks in the $J_\mathrm{T}(q, \omega)$ we can derive an
associated transverse dispersion relation, $\omega_\mathrm{T}(q)$, which is plotted in
figure~\ref{disperT}.
The $\omega_\mathrm{T}(q)$ shows, for small $q$-values, a linear behaviour which
can be approximated by $\omega_\mathrm{T}(q)\approx c_{\mathrm{t}} (q-q_c)$, where $q_c$ is the value at which
the $J_\mathrm{T}(q, \omega)$ starts showing a maximum and $c_{\mathrm{t}}$ is the velocity of propagation of
the shear waves in the liquid metal. In the case of l--Cu at 1473 K, we obtained
an estimate $c_{\mathrm{t}}\approx 3000  \pm 200$~m/s.

From the previous $J_\mathrm{T}(q,t)$ and using the
above mentioned procedure, we evaluated the
shear viscosity for l---Cu. This was performed by calculating the
generalized  shear viscosity coefficient $\widetilde{\eta}(q)$ for a
range $q \leqslant 0.8$~\AA$^{-1}$ and fitting it to the expression
$\widetilde{\eta}(q)=\eta(1-\alpha\,q^{2})$. Herein we
estimated $\eta \approx$
$3.63\cdot  10^{-3}$~kg/ms (for $T=1423$~K) and $2.25\cdot  10^{-3}$~kg/ms ($T= 1773$~K)
which compare rather well with the corresponding experimental
data of $3.98\cdot  10^{-3}$~kg/ms
and $2.39\cdot  10^{-3}$~kg/ms, respectively~\cite{ShimojiBook2}.

\subsubsection{Liquid Ag}

\begin{figure}[!h]
\centerline{
\includegraphics[width=0.44\textwidth,clip]{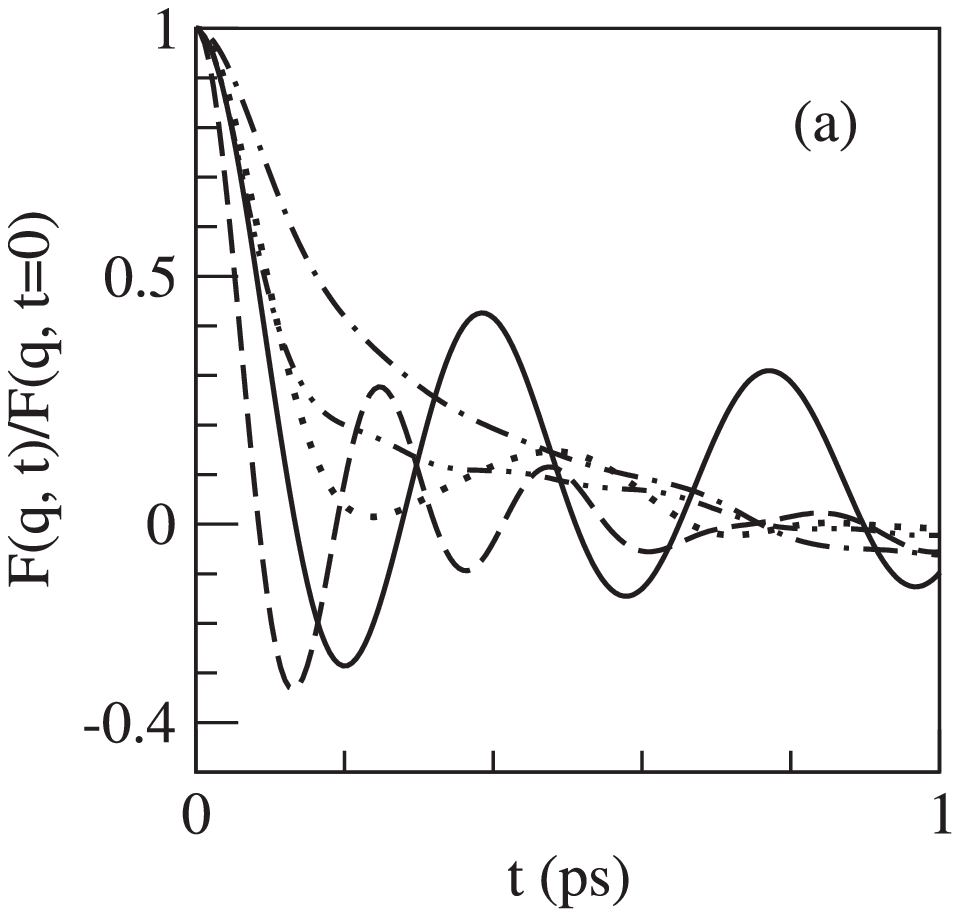}\hspace{8mm}
\includegraphics[width=0.44\textwidth,clip]{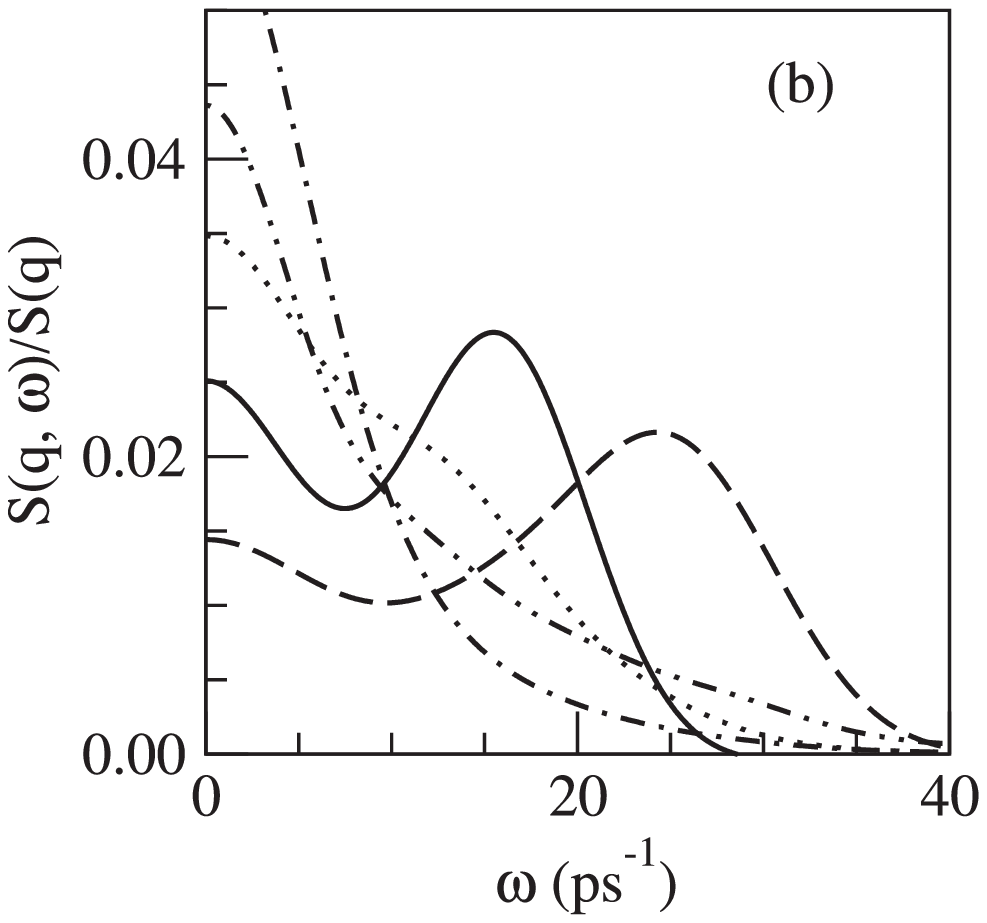}
}
\caption{(a) Intermediate scattering function of
l--Ag at $T=1273$~K.
Full line: 0.59~\AA$^{-1}$,
dashed line: 1.5~\AA$^{-1}$,
dotted line: 2.1~\AA$^{-1}$,
dotted-dashed line: 3.0~\AA$^{-1}$,
and double dotted-dashed line: 4.2~\AA$^{-1}$
(b) Same as above but for the dynamic structure factor.}
\label{Fqt-Ag}
\end{figure}

Figure~\ref{Fqt-Ag}~(a) shows the calculated $F(q,t)/F(q,t=0)$
at several $q$-values for l--Ag at $T= 1273$~K.
Again,
for small $q$-values we observe an oscillatory behaviour which
is gradually dampened with increasing $q$-values until it finally disappears
at $q \approx 0.75 q_\mathrm{p}$.
The corresponding $S(q,\omega)$ are plotted in
figure~\ref{Fqt-Ag}~(b)
and we observe side-peaks for a range of small $q$-values, namely up to
$q \approx 0.75 q_\mathrm{p}$.
From the position of the
peaks of $J_\mathrm{L}(q,\omega)$ we obtain the corresponding dispersion relation, $\omega_\mathrm{L}(q)$,
plotted for $T=1273$~K in figure~\ref{disperL}.
Using the smallest $q$-value provided due to the simulations,
$q_\mathrm{min}= 0.294$~\AA$^{-1}$,
we get
an estimate $c_{\mathrm{s}} (q_\mathrm{min}) \approx$ 2930 m/s, which is somewhat greater than
the hydrodynamic values of $c_{\mathrm{s}} = 2751$~\cite{Blairs},
2710~\cite{BeyerRing} or 2797 m/s~\cite{Singh},
and suggests the existence of a small positive dispersion effect.
At a higher temperature, $T=1673$~K, our calculation predicts a value  of
2720 m/s, whereas the experimental adiabatic sound velocity is 2663 m/s~\cite{Singh}.
We are not aware of any inelastic XR or neutron scattering experiments for liquid Ag
to compare with.

\begin{figure}[!h]
\centerline{
\includegraphics[width=0.44\textwidth,clip]{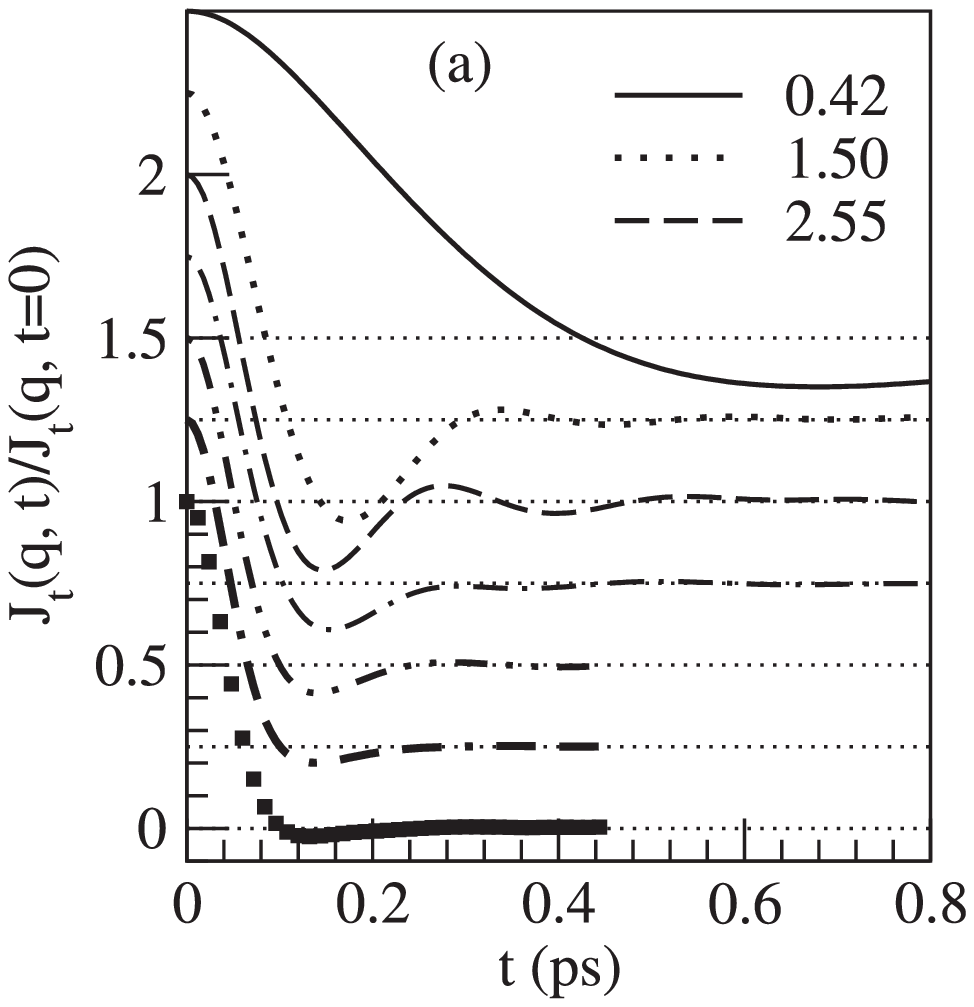}\hspace{8mm}
\includegraphics[width=0.44\textwidth,clip]{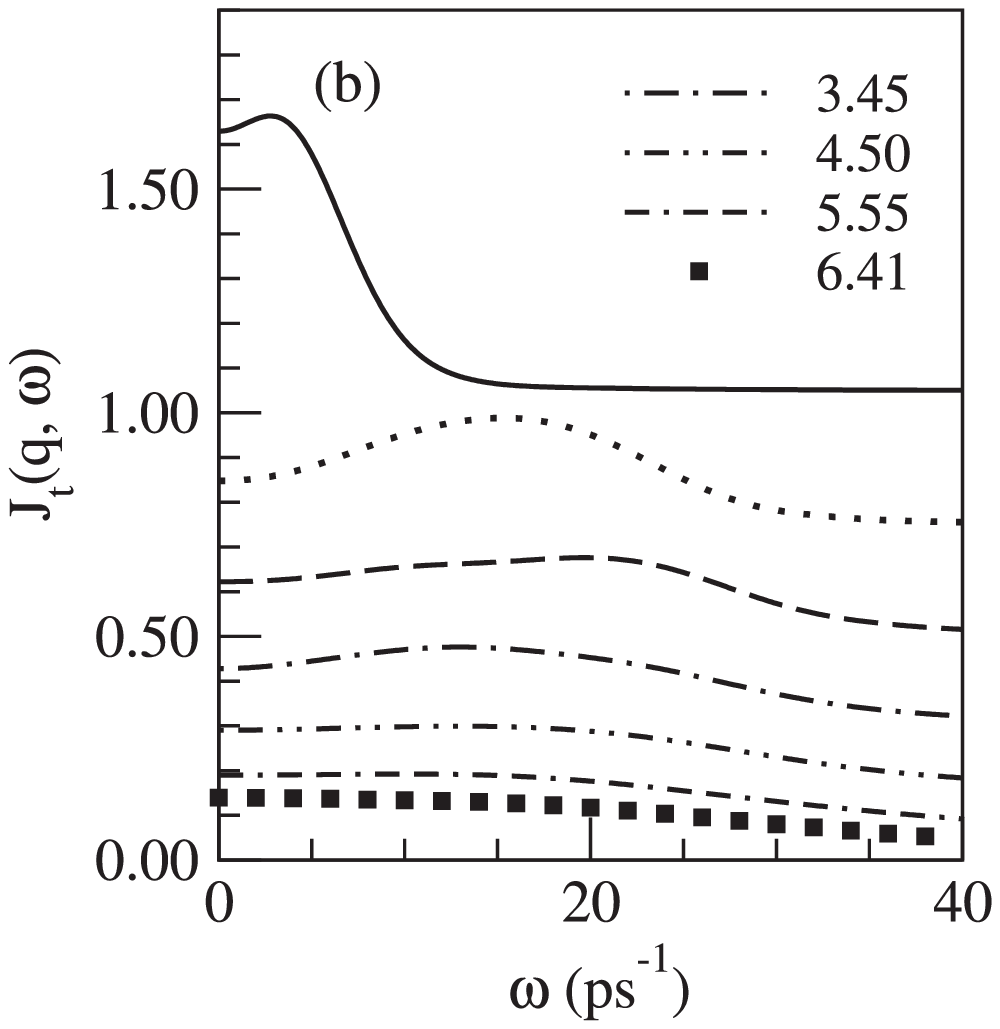}
}
\caption{(a) Transverse current correlation function of
l--Ag at $T=1273$~K.
(b) Same as above but for the $J_\mathrm{T}(q, \omega)$.}
\label{Trans-Ag}
\end{figure}

Figure~\ref{Trans-Ag} shows the
calculated $J_\mathrm{T}(q,t)$ and their Fourier Transforms
for l--Ag at $T=1423$~K and
for several $q$ values. The main features for
$J_\mathrm{T}(q,t)$ and  $J_\mathrm{T}(q,\omega)$ are similar to those found in l--Cu.
The $J_\mathrm{T}(q,\omega)$ shows the peaks from which the corresponding dispersion
relation $\omega_\mathrm{T}(q)$ was
calculated and it is depicted in figure~\ref{disperT}. Again, a linear fit of the small $q$
values yields an estimate of $c_{\mathrm{t}} \approx 1950 \pm 150$~m/s for the velocity of the
associated shear waves.
The calculation of the shear viscosity  yields
$\eta=3.48\cdot 10^{-3}$~kg/ms and $2.16\cdot 10^{-3}$
at $T=1273$~K and $1673$~K, which is in good agreement with the
respective  experimental
data $\eta=3.69\cdot 10^{-3}$ and
$2.23\cdot 10^{-3}$~kg/ms ~\cite{ShimojiBook2}.

\subsubsection{Liquid  Au}

\begin{figure}[!t]
\centerline{
\includegraphics[width=0.44\textwidth,clip]{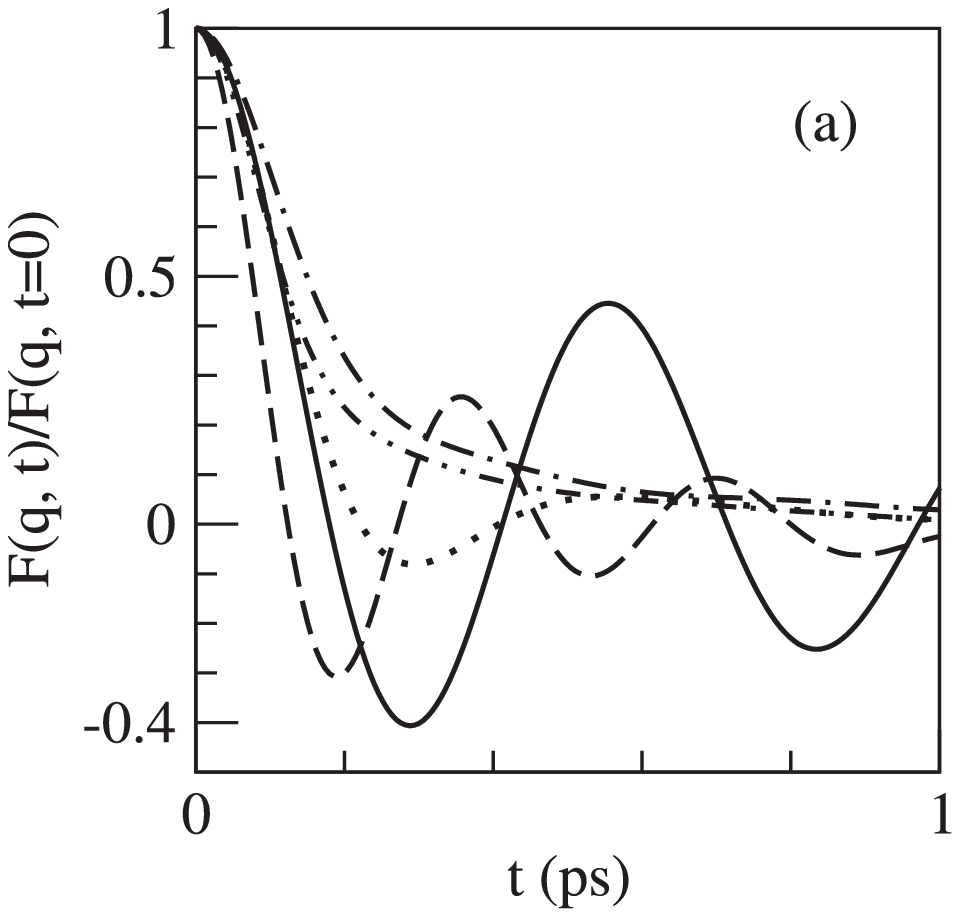}\hspace{8mm}
\includegraphics[width=0.44\textwidth,clip]{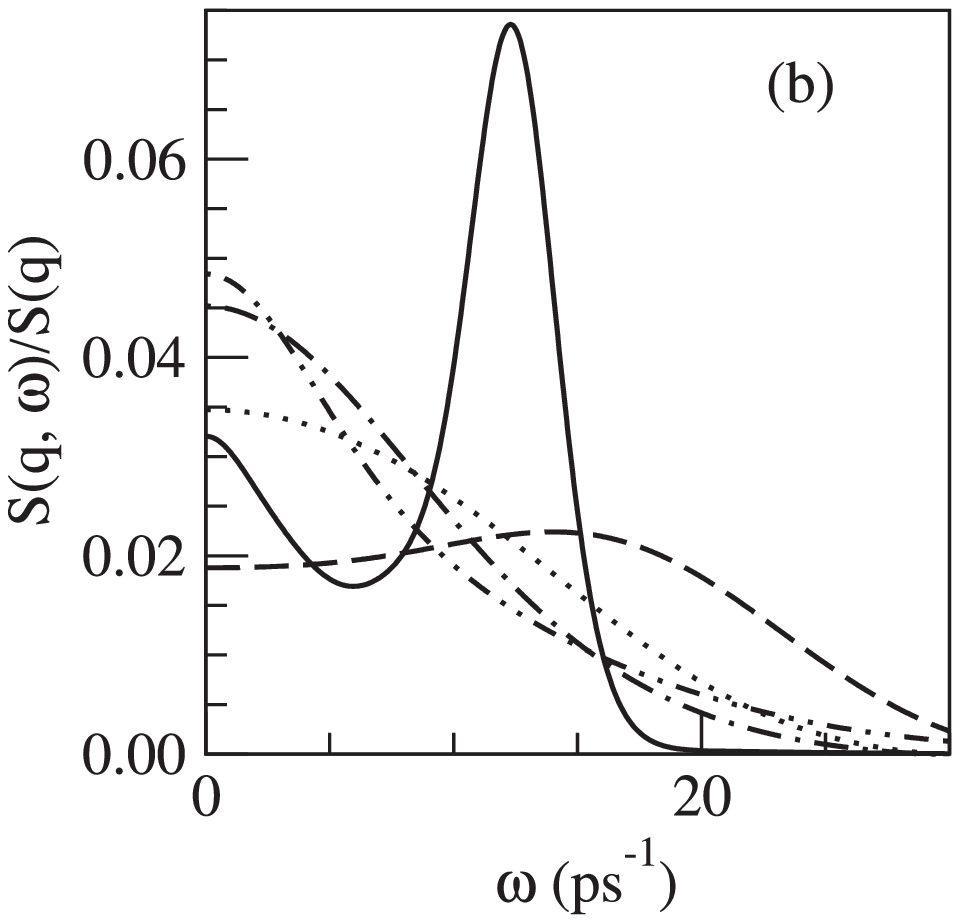}
}
\caption{(a) Intermediate scattering function of
l--Au at $T=1423$~K. Full line: 0.59~\AA$^{-1}$, dashed line: 1.39~\AA$^{-1}$,
dotted line: 2.0~\AA$^{-1}$, dot-dashed line: 3.2~\AA$^{-1}$
and double dot-dashed line: 4.2~\AA$^{-1}$
(b) Same as above but for the dynamic structure factor.}
\label{Fqt-Au}
\end{figure}

\begin{figure}[!b]
\centerline{
\includegraphics[width=0.422\textwidth,clip]{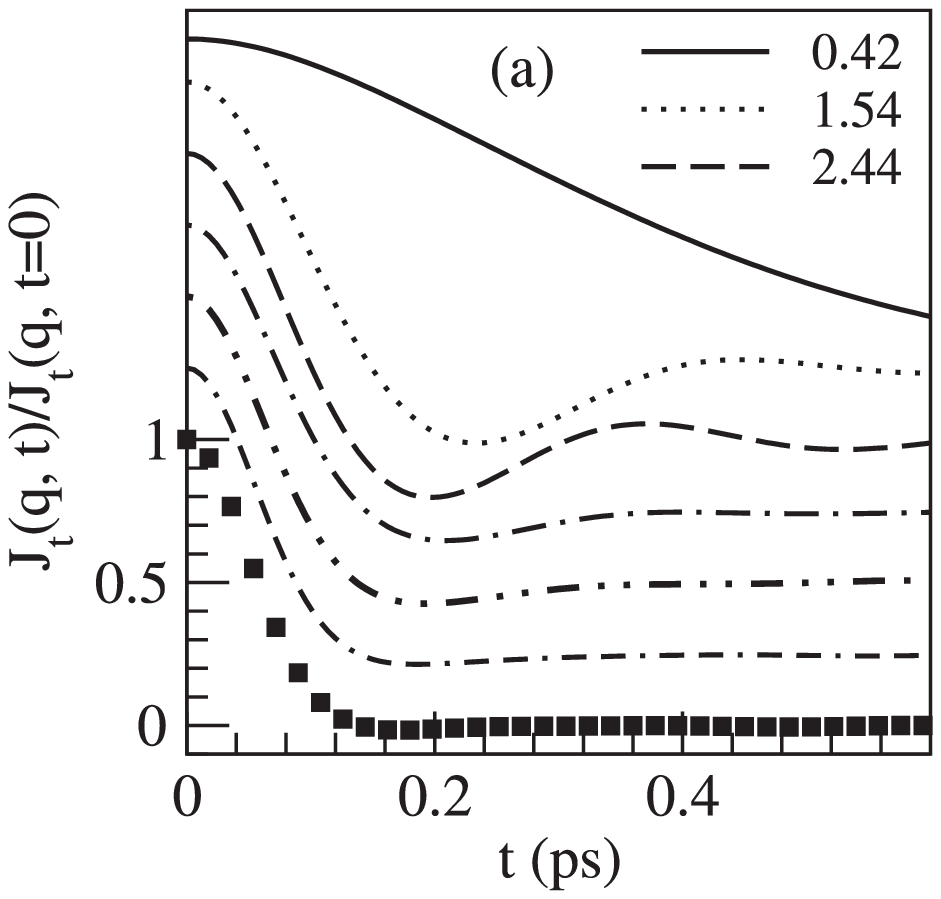}\hspace{8mm}
\includegraphics[width=0.44\textwidth,clip]{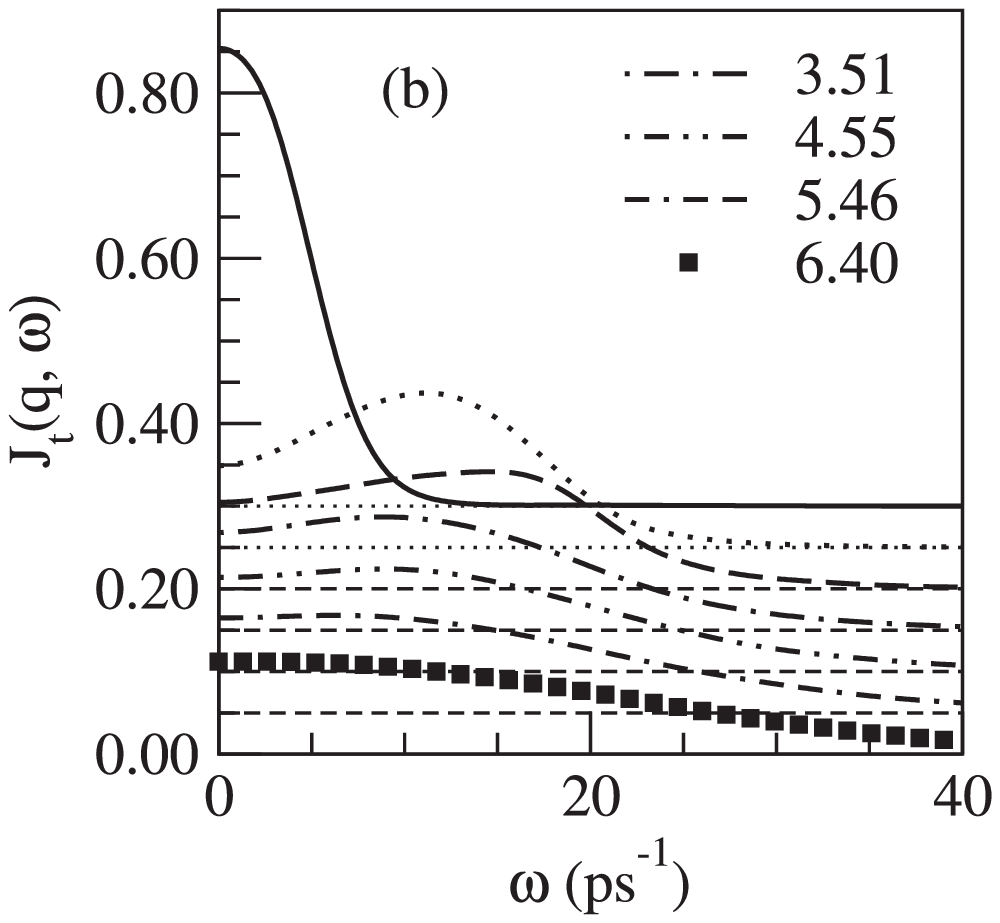}
}
\caption{(a) Transverse current correlation function of
l--Au at $T=1423$~K.
(b) Same as (a) but for the $J_\mathrm{T}(q, \omega)$.}
\label{Trans-Au}
\end{figure}

Figure~\ref{Fqt-Au}~(a) shows the calculated $F(q,t)/F(q,t=0)$
for several $q$-values. Their main features are similar to those found in
l--Cu and l--Ag, namely the  existence of oscillations up to $q \approx 0.75 q_\mathrm{p}$.
Figure~\ref{Fqt-Au}~(b) depicts, for several $q$-values, the corresponding
$S(q,\omega)$ which show clear side-peaks up to $q \approx 0.75 q_\mathrm{p}$.
The dispersion relation of the longitudinal currents is plotted in
figure~\ref{disperL} for $T=1423$~K.
From the  smallest $q$-value in the simulations,
$q_\mathrm{min}= 0.296 $~\AA$^{-1}$,  we get
an estimate $c_{\mathrm{s}} (q_\mathrm{min}) \approx$ 2030 m/s, which is clearly below
the hydrodynamic adiabatic value of $c_{\mathrm{s}} =$ 2567 m/s for 1337 K~\cite{IGBook,Blairs}
and 2513 m/s at 1423 K~\cite{Singh}. This points towards the presence of negative dispersion
in the dispersion relation.
Although negative dispersion has been indeed found and explained in some systems,
such as supercritical fluids~\cite{Bryk-JCP}, as a consequence of an increased ratio between
the high-frequency sound velocity and the adiabatic one, driven mainly by a decreased
value of the density, the present case of liquid Au near melting certainly does not fit
into this category of liquids.
Two scenarios could possibly explain the negative dispersion obtained in our calculation.
The first one is related to the value of the isothermal speed of sound, $c_{\mathrm{T}}$.
Blairs' data~\cite{Blairs} are consistent with a value of $\gamma=1.50$ at 1337~K, which
yields a value of $c_{\mathrm{T}}= 2096$~m/s, so that at 1423~K, the isothermal sound
velocity would be somewhat smaller, i.e., quite similar to our result for $c_{\mathrm{s}}(q_\mathrm{min})$.
Singh {et al.}'s data~\cite{Singh} give, however, a value of $\gamma=1.36$ at 1336~K, and
$\gamma=1.40$ at 1423~K, producing an isothermal sound velocity of 2124~m/s at 1423~K,
which is also similar, although still larger, than our $c_{\mathrm{s}}(q_\mathrm{min})$
We could, therefore, argue that we are in a wavenumber domain where sound propagation
is isothermal in nature. This effect was indeed
found in l-Ni~\cite{Scopigno,Bermejo} and it is connected with the existence of
an intermediate isothermal
domain standing between hydrodynamic and high-frequency domains~\cite{ScoRuo}.
The second scenario is simply a scenario of either inaccuracies in the theoretical method that
lead to a {\em simulation} adiabatic sound velocity which is too small compared with experiment,
or inaccuracies in the experimental data which would report too high a value of the
real sound velocity. In either case, the negative dispersion that we find would just be
an artifact produced by the wrong value of the hydrodynamic sound velocity.
In this respect, it is worth recalling that just one measurement
of the speed of sound in liquid Au exists~\cite{Au-sound}. It is also worth mentioning that
very few theoretical calculations of this property can be found in the literature.
For instance, the only KS-AIMD simulations of liquid Au~\cite{Jakse-Au} did  not address
the collective dynamics. Concerning CMD, we have only found one reference where a value of
$c_{\mathrm{s}}$ is mentioned~\cite{Ercolessi-Au}, which was obtained using a glue model interatomic
potential, successfully used previously to study several properties of solid Au; the
value of $c_{\mathrm{s}}$ was 3700 m/s at 1360 K, which is notoriously high as compared to our result and
the experimental value.

\begin{figure}[!t]
\begin{center}
\includegraphics[width=0.44\textwidth,clip]{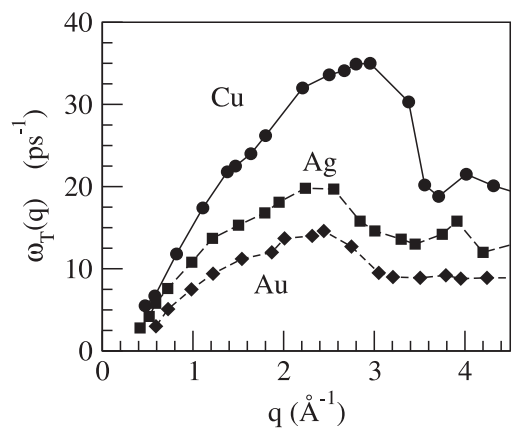}
\caption{Transverse dispersion relations
for l--Cu, l--Ag and l--Au at T = 1423, 1273  and 1423~K, respectively.}
\label{disperT}
\end{center}
\end{figure}

Figure~\ref{Trans-Au} shows the calculated
$J_\mathrm{T}(q,t)$ and $J_\mathrm{T}(q,\omega)$
for l--Au at $T=1423$~K and several $q$ values.
The corresponding transverse dispersion relation is plotted in figure~\ref{disperT}.
Its low $q$ behaviour leads to an estimate
$c_{\mathrm{t}}\approx 1380 \pm 150$~m/s for the velocity of the
corresponding shear waves.
The calculation of the shear viscosity has yielded values
$\approx 4.05 \cdot  10^{-3}$~kg/ms and $3.304\cdot  10^{-3}$~kg/ms
for  $T=1423$~K and $1773$~K, respectively; these are close to the
corresponding experimental data of $4.34\cdot  10^{-3}$~kg/ms
and \linebreak $3.33
\cdot  10^{-3}$~kg/ms~\cite{ShimojiBook2}.

\section{Conclusion}

We have reported several static and dynamic properties of the liquid noble
metals (Cu, Ag, Au), each at two thermodynamic states near their respective
triple points. This was carried out by using the orbital free
{\it ab-initio} molecular dynamic simulation method which has
already shown its capability for yielding accurate estimates of the same
properties for a range
of simple metals and alloys~\cite{GGLS,GGLS_2}.

The static structure of the three systems at the two temperatures is globally
very well described. Only the low-$q$ part of the structure factor differs from
the values obtained from thermodynamic data in l--Cu, where $S(0)$ is underestimated
and in l--Au where it is overestimated.
The close similarities between the structures of l--Ag and l--Au near melting are indeed
reproduced within our model. For both systems, the main peaks in their
respective $g(r)$ and $S(q)$ are  located at very similar positions. This can be explained
by noting that the static structure is mostly determined by the repulsive
part of the interionic interaction and density of ions. Figure~\ref{psfig}
shows that the
repulsive part of the non-Coulombic part of the electron ion interaction for l--Ag
and l--Au are practically coincidental whereas the experimental ionic number densities
near melting are 0.0551~\AA$^{-3}$ (l--Ag) and 0.0525~\AA$^{-3}$ (l--Au).

As for the dynamic properties, we begin by noting that the calculated $Z(t)$ show
the characteristic shape of high density systems~\cite{GGLS,GGLS_2,Balubook}
(i.e., the simple liquid metals near melting), which can be explained in terms of
the so-called cage effect, namely, a tagged particle is enclosed in a cage
formed by its adjacent neighbors.  Results have also been reported for the
selfdiffusion coefficients, adiabatic sound velocities and shear viscosities.
The calculated dynamic structure factors, $S(q, \omega)$, show side-peaks up
to $q \approx 0.75 q_\mathrm{p}$, which is similar, albeit a bit larger than that of the
simple liquid metals~\cite{GGLS,GGLS_2,Balubook}. The calculated dispersion relations suggest the
existence of some positive dispersion in l--Cu and, to a smaller extent, in l--Ag;
in the case of l--Au, some negative dispersion appears to happen, but we could not ellucidate
whether it is a real feature or it is due to inaccuracies in the experimental data
or in the theoretical model.

We conclude, as far as the agreement with the available experimental data is
concerned the present OF-AIMD results for static and dynamic properties are very good.
Most importantly, the results also show the capability and reliability of
our approach in handling very complicated $d$-electron systems in liquid phase
from the perspective of {\em ab-initio} studies. Additional OF-AIMD calculations combined with
our model for the pseudopotential are already in progress for several liquid transition
metals. Preliminary results are very encouraging and those will be reported in due course.

\section*{Acknowledgements}

This work was supported by MICINN in conjunction with the EU FEDER funds
(project FIS2011--22957) and by Junta de Castilla y Leon (project VA104A11--2).
DJG additionally acknowledges financial support from
MECD (project PR2011--0019).
GMB is grateful to the Universidad de Valladolid for a
fellowship which allowed him to carry out this work.
He also gratefully acknowledges the hospitality provided by
Professor D. J. Gonz\'alez and L. E. Gonz\'alez during his stay at the
Universidad de Valladolid.

\newpage
\ukrainianpart

\title{Дослідження деяких статичних та динамічних властивостей рідких благородних металів методом безорбітальної першопринципної молекулярної динаміки}
\author{Ґ.М. Бгуян\refaddr{label1}, Л.Е. Ґонсалес\refaddr{label2}, Д.Дж. Ґонсалесалес\refaddr{label2}}

\addresses{
\addr{label1} Факультет теоретичної фізики, університет м. Дака, Дака--1000, Бангладеш
\addr{label2} Факультет теоретичної фізики, університет м. Вальядолід, Вальядолід, Іспанія
}

\makeukrtitle

\begin{abstract}
\tolerance=3000%
Оцінено деякі статичні та динамічні властивості рідких  Cu, Ag і Au при термодинамічних умовах, близьких до їх точок плавлення, за допомогою методу безорбітальної першопринципної молекулярної динаміки. Розрахована статична структура  добре узгоджується  з наявними даними, отриманими з  рентгенівської і нейтронної дифракцій. Щодо динамічних властивостей, то розраховані динамічні структурні фактори вказують на існування колективних збуджень густини  поряд з позитивною дисперсією для
l--Cu і l--Ag. Отримано  деякі коефіцієнти переносу, які прийнятно узгоджуються з наявними експериментальними даними.

\keywords рідкі благородні метали, безорбітальна теорія функціоналу густини, моделювання методом молекулярної динаміки, статична структура, динамічні властивості, коефіцієнти переносу
\end{abstract}

\end{document}